\documentclass{article}
\usepackage[utf8]{inputenc}
\usepackage[english]{babel}
\usepackage{xcolor}
\usepackage{hyperref}
\usepackage{cite}  
\usepackage{algorithm} 
\usepackage{algpseudocode} 
\usepackage{graphicx}
\usepackage{amsmath}
\usepackage{bm}
\usepackage{soul}
\usepackage{authblk}

\algnewcommand\algorithmicswitch{\textbf{switch}}
\algnewcommand\algorithmiccase{\textbf{case}}
\algnewcommand\algorithmicassert{\texttt{assert}}
\algnewcommand\Assert[1]{\State \algorithmicassert(#1)}%
\algdef{SE}[SWITCH]{Switch}{EndSwitch}[1]{\algorithmicswitch\ #1\ \algorithmicdo}{\algorithmicend\ \algorithmicswitch}%
\algdef{SE}[CASE]{Case}{EndCase}[1]{\algorithmiccase\ #1}{\algorithmicend\ \algorithmiccase}%
\algtext*{EndSwitch}%
\algtext*{EndCase}%

\title{Polyhedron Kernel Computation Using a Geometric Approach}
\author{T.~Sorgente}
\author{S.~Biasotti}
\author{M.~Spagnuolo}
\affil{Istituto di Matematica Applicata e Tecnologie Informatiche `E. Magenes' - CNR, Italy}

\begin{document}
\maketitle

\begin{abstract}
   The geometric kernel (or simply the kernel) of a polyhedron is the set of points from which the whole polyhedron is visible. 
   Whilst the computation of the kernel for a polygon has been largely addressed in the literature, fewer methods have been proposed for polyhedra.
   The most acknowledged solution for the kernel estimation is to solve a linear programming problem. 
   On the contrary, we present a geometric approach that extends our previous method \cite{SorgenteKernel}, optimizes it anticipating all calculations in a pre-processing step and introduces the use of geometric exact predicates.
   Experimental results show that our method is more efficient than the algebraic approach on generic tessellations and in detecting if a polyhedron is not star-shaped.  
   Details on the technical implementation and discussions on pros and cons of the method are also provided.
\end{abstract}


\section{Introduction}
\label{sec:intro}

\begin{figure*}
    \includegraphics[width=\linewidth]{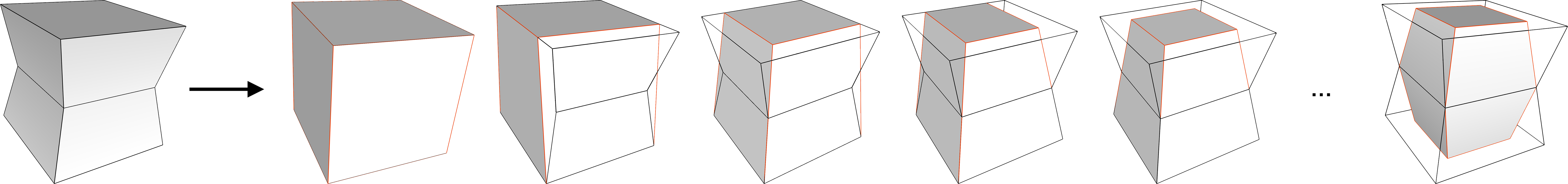}
    \centering
    \caption{Pipeline of the kernel computation for a polyhedron. At first step, we compute the Axis Aligned Bounding Box (AABB) of the polyhedron; then, we iterate on each face $f$ of the polyhedron (black edges) and cut AABB with the plane induced by $f$ (red edges).}
    \label{fig:pipeline}
\end{figure*}

The concept of \emph{geometric kernel} of a polygon, a polyhedron, or more generally of a shape, is a pillar of computational geometry. 
Roughly speaking, the kernel of a closed geometric shape $S$ is the locus of the points internal to $S$ from which the whole shape $S$ is visible.
The kernel of a convex shape coincides with the shape itself, while the concept is particularly interesting for non-convex shapes and in particular, polytopes. The kernel of non-convex shapes can also be empty, as in the case of non simply-connected objects for which it is always empty.

In the two-dimensional scenario, that is when the shape is a polygon, the standard way of computing the kernel is by intersecting appropriate half-planes generated from its edges.
This problem has been tackled since the 70s, when \cite{ShamosHoey} presented an algorithm able to perform the kernel computation in $O(e\log e)$ operations, being $e$ the number of edges of a polygon, as the intersection of $e$ half-edges.
After that, an optimal algorithm able to run in $O(n)$ operations over an $n-$sided polygon, has been proposed in \cite{LeePreparata}. 
Famous computational tools and libraries like \textit{Boost} \cite{BoostLibrary}, \textit{Geogram} \cite{levy2015geogram}, \textit{CGAL} \cite{fabri2009cgal}, or \textit{Libigl} \cite{jacobson2017libigl} currently implement routines to compute intersections between polygons and planes, which can be used to estimate the kernel.
In the first attempts to solve the volumetric version of the problem, for example in \cite{PreparataShamos}, the natural approach has been to extend the 2D method (which we call \textit{geometric)} to the 3D case from a theoretical point of view, but it was soon dismissed as unattractive for computational reasons.
It was replaced by a new approach (which we call \textit{algebraic)} which makes use of linear algebra and homogeneous coordinates, and that is the state of the art for computing 3D kernels currently implemented by libraries like CGAL.

During years, the polygon kernel computation has become popular to address several problems based on simple polygon analysis, such as star-components decomposition and visibility algorithms that are of interest in robotics, surveillance, geometric modeling, computer vision and, recently, in the emerging field of additive manufacturing \cite{demir2018near}.

Today, the geometric kernel of a polytope is a pivotal information for understanding the geometrical quality of an element in the context of finite elements analysis.
While in the past years finite elements methods were only designed to work on convex elements like triangles/tetrahedra or quadrangles/hexahedra  \cite{ciarlet2002finite}, recent and more complex methods like the Mimetic Finite Difference Method \cite{lipnikov2014mimetic}, the Virtual Elements Method \cite{beirao2013basic}, the Discontinuous Galerkin Mehod \cite{cockburn2012discontinuous} or the Hybrid High Order Method \cite{di2019hybrid} are able to deal with non convex polytopes.
This enrichment of the class of admissible elements led researchers to further investigate the concept of the geometric quality of a polytope, and to define quality measures and metrics for the mesh elements, whether they are poligons \cite{attene2021benchmark,sorgente2022role} or polyhedra \cite{sorgente2021polyhedral}.
In this setting, the geometric kernel is often associated with the concepts of \textit{shape regularity} and \textit{star-shapedness} of an element.
For example, as analyzed in \cite{sorgente2021vem}, most of the error estimates regarding the Virtual Elements Method (but the same holds for other polytopal methods) are based on the theory of polynomial approximation in Sobolev spaces, assuming the star-shapedness of the elements \cite{Brenner-Scott:2008,dupont1980polynomial}.
As a consequence there are a number of sufficient geometrical assumptions on the computational domain for the convergence of the method, which require an estimate of the kernel. 
When dealing with non-trivial meshes \cite{antonietti2022high, berrone2019parallel}, such quality measures/metrics/indicators require to compute the kernel of thousands of polytopes, each of them with a limited number of faces and vertices, in the shortest possible time.

A preliminary algorithm for the computation of the kernel of a polyhedron using the geometric approach was introduced in \cite{SorgenteKernel}. 
There, we experimentally showed how this type of approach in practice can significantly outperform the algebraic one, for instance, when polyhedral elements have a limited number of faces and vertices or several faces are co-planar. 
In this paper we optimize the algorithm, for instance preliminary estimating the position of all vertices with respect to the planes induced by the faces, introducing the use of exact geometric predicates \cite{shewchuk1997adaptive} and a random strategy for visiting the faces of a polyhedron and identify more rapidly polyhedra whose kernel is empty. 
We also simplify the description of the algorithmic routines and we introduce new scientific results on a larger variety of models that confirm the validity of a geometry-based kernel algorithm.

The paper is organized as follows.
In Section~\ref{sec:preliminary} we introduce some notation and discuss the difference between the geometric and the algebraic approach.
In Section~\ref{sec:kernel} we detail the algorithm for the construction of the kernel of a polyhedron.
In Section~\ref{sec:tests} we exhibit some examples of computed kernels and analyze the performance of the algorithm with comparisons with an implementation of the algebraic approach and with its previous version \cite{SorgenteKernel}.
In Section~\ref{sec:conclusions} we sum up pros and cons of the algorithm and draw some conclusions.

\section{Preliminary concepts}
\label{sec:preliminary}
We introduce some notations and preliminary concepts that we will use in the rest of the paper.


\subsection{Notations}
\label{subsec:notations}
We define a \textit{polyhedron} as a finite set of plane polygons such that every edge of a polygon is shared by exactly one other polygon and no subset of polygons has the same property \cite{PreparataShamos}.
The vertices and the edges of the polygons are the \textit{vertices} and the \textit{edges} of the polyhedron; the polygons are the \textit{faces} of the polyhedron.
In this work we only consider \textit{simple} polyhedra, which means that there is no pair of nonadjacent faces sharing a point.

A polyhedron $P$ is said to be \textit{convex} if, given any two points $p_1$ and $p_2$ in $P$, the line segment connecting $p_1$ and $p_2$ is entirely contained in $P$.
It can be shown that the intersection of convex polyhedra is a convex polyhedron \cite{PreparataShamos}.
Two points $p_1$ and $p_2$ in $P$ are said to be \textit{visible} from each other if the segment $(p_1,p_2)$ does not intersect the boundary of $P$. 
The \textit{kernel} of $P$ is the set of points in the interior of $P$ from which all the points in $P$ are visible. 
The first obvious consideration is that the kernel of a polyhedron is a convex polyhedron.
If $P$ is convex its kernel coincides with its interior, because any two points inside a convex polyhedron are visible from each other.
A polyhedron may also not have a kernel at all; in this case we say that its kernel is \textit{empty}.
Last, a polyhedron $P$ is called \textit{star-shaped} if there exists a sphere, with non-zero radius, completely contained in its kernel.
A polyhedron is star-shaped if and only if its kernel is not empty, therefore star-shapedness can be thought as an indicator of the existence of a kernel.
Star-shapedness is weaker than convexity, and it is often used in the literature as many theoretical results in the theory of polynomial approximation in Sobolev spaces rely on this condition \cite{dupont1980polynomial, Brenner-Scott:2008}.

\begin{figure}[htbp]
\centering
    \includegraphics[width=\linewidth]{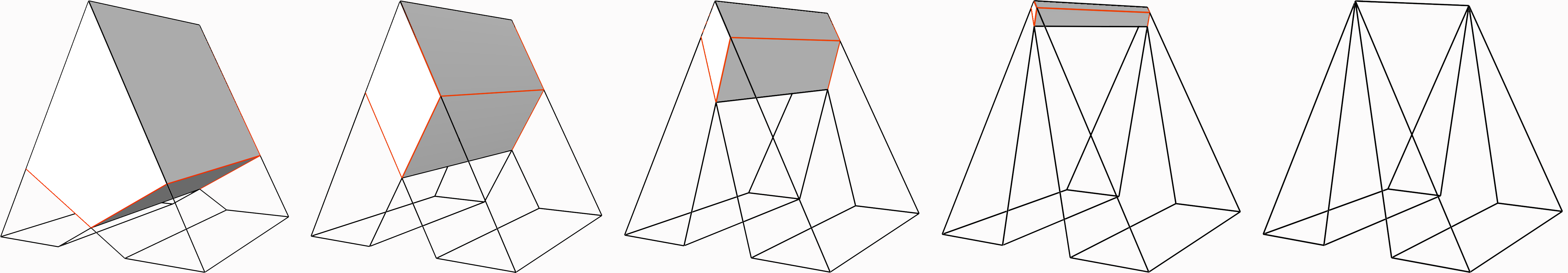}
\caption{A sequence of parametric polyhedra whose kernel is progressively shrinking. The kernel is the polyhedron delimited by the red edges.}
\label{fig:tent}
\end{figure}

In order to give a visual example of these concepts, in Fig.~\ref{fig:tent} we present a parametric object shaped like a tent, with the parameter regulating the height of the ``entrance'' from the basis.
This object is not convex, but in the first examples it is star-shaped and it has a proper kernel, delimited by the red edges.
As the parameter increases, the set of points from which the whole polyhedron is visible becomes smaller, and so does the star-shapedness radius.
The last example of Fig.~\ref{fig:tent} is not star-shaped anymore, i.e. the kernel is empty.


\subsection{The algebraic approach to the kernel computation}
As observed in Section~\ref{sec:intro}, the state of the art algorithm for the computation of the kernel of a polygon follows a geometric approach: the kernel is found as the intersection of half-planes originating from its edges.
We use the term ``geometrical'' because the algorithm computes repeatedly a sequence of geometric intersections between polygons and planes.
This idea was afterwards optimized until obtaining an algorithm able to run in $O(n)$ operations, which has been proven to be optimal \cite{LeePreparata}.

When facing the 3D version of the problem, one natural way could be to extend the 2D algorithm, which is well studied and documented, to the higher dimension.
The problem with the 3D case is that, whereas two convex polygons with respectively $n_1$ and $n_2$ vertices can be intersected in time $O(n)$, being $n=n_1+n_2$, two convex polyhedra with the same parameters are intersected in time $O(n\log n)$, thus the generalization of the two-dimensional instance would yield an $O(n\log^2 n)$ algorithm.
This is in contrast with the result shown in \cite{PreparataShamos}, where a lower bound for the intersection of convex polyhedra is established at $O(n\log n)$.
Therefore the geometric approach to the 3D problem was soon dismissed as unattractive, and alternative ways have been explored.

A new algorithm was formulated, based on the so-called ``double duality trick'', which makes use of linear algebra and homogeneous coordinates.
Thanks to this algebraic approach, described in \cite[Section 7.3.2]{PreparataShamos}, it is possible to compute the intersection of $n$ half-spaces in time $O(n\log n)$ \cite{PreparataShamos}.
This algorithm can be implemented inside the framework of the CGAL library, although there is currently not an explicit routine for computing the kernel of a polyhedron and one has to connect the function for the intersection of half-spaces to the polyhedron data structure.
The whole routine implies:
\begin{enumerate}
    \item solving a linear problem to find a point in the interior of the polyhedron;
    \item translating the polyhedron so that the interior point is the origin;
    \item calculating the dual polyhedron, defined as the convex hull of the vertices dual to the original faces in regard to the unit sphere (i.e., halfspaces at distance $d$ from the origin are dual to vertices at distance $1/d$);
    \item calculating the resulting polyhedron, which is the dual of the dual polyhedron;
    \item translating the origin back to the interior point.
\end{enumerate}

While from a theoretical point of view the cited results are indubitable, we believe that in many practical situations the geometric approach could perform better than the algebraic one.
This is due to the fact that the cost of solving a linear problem cannot fall below a certain bound, while the intersection of half-spaces can become extremely cheap in some circumstances.
For instance if the number of faces and vertices of the polyhedron is low, this method could be more efficient than the algebraic one, and there may be other situations in which one could take advantage of a wise treatment of the geometrical operations.


\subsection{Data structure}
\label{subsec:data}
We adopt the following data structure inherited by the \textit{cinolib} library \cite{livesu2019cinolib}, in which the code has been written:
\begin{itemize}
    \item \textit{Points:} array of unordered 3D points.
    \item \textit{Face:} array of unsigned integers associated to a \textit{Points} array, representing the indices of the vertices of a face ordered counter-clockwise.
    \item \textit{Polyhedron:} struct composed by a field \textit{verts} of type \textit{Points} containing the vertices and a field \textit{faces} of type \textit{array
    $<$Face$>$} containing the faces of a polyhedron.
    \item \textit{Plane:} class defining a plane in the Hessian form, composed by a 3D point $n$ indicating the unit normal of the plane (i.e. the $a,b,c$ coefficients of the plane equation) and a number $d$ indicating the distance of the plane from the origin (i.e. the $d$ coefficient of the plane equation).
    The plane class also contains three additional points $p_1,p_2,p_3$ lying on the plane, useful for the Shewchuck exact predicates \cite{shewchuk1997adaptive}.
    \item \textit{Sign:} array of labels (BELOW, ABOVE or INTER) used to store information on the position of the elements of a polyhedron (points, edges or faces) with respect to an unspecified plane.
\end{itemize}
Given a plane $p$, the elements of a polyhedron are classified as follows:
\begin{itemize}
    \item A point $v$ is labelled as BELOW, ABOVE or INTER provided that the function \textit{orient3d}$(p.p_1,p.p_2,p.p_3,v)$ in the Shewchuck exact predicate library \cite{shewchuk1997adaptive} is negative, positive or zero, within a tolerance of $10^{-8}$;
    \item An edge $e$ is labelled as BELOW (resp. ABOVE) if both its endpoints are BELOW or INTER (resp. ABOVE or INTER), and as INTER if one point is ABOVE and the other one is BELOW.
    \item A face $f$ is labelled as BELOW if all of its points are BELOW, as ABOVE if none of its points is BELOW and as INTER otherwise.
\end{itemize}
The classification of points is computed at the top level of the algorithm with Shewchuck exact predicates, defining the symbol CINOLIB$\_$USES$\_$EXACT$\_$PREDICATES at compilation time.
For edges and faces instead, we define a function \textit{classify(S)} which determines the classification of and edge or a face from an array $S$ of type \textit{Sign} containing the classification of its points.
The introduction of the labels and the points evaluation at the beginning of the process are two main computational novelties with respect to the previous version of the algorithm \cite{SorgenteKernel}.

\section{Polyhedron kernel algorithm}
\label{sec:kernel}
In this section, we illustrate our method for computing the kernel of a polyhedron with a geometric approach.
It has a modular structure composed of four nested algorithms, each one calling the next one in its core part.
It is modular in the sense that each algorithm can be entirely replaced by another one performing the same operation(s).
This property is particularly useful for making comparisons: one could, for instance, use different strategies for computing the intersection between a polygon and a plane and simply replace Algorithm~\ref{alg:polygon-plane}, measuring the efficiency from time to time.


\subsection{Polyhedron Kernel}
\label{subsec:polyhedron_kernel}
With Algorithm~\ref{alg:kernel} we tackle the general problem: given a polyhedron $P$, we want to find the polyhedron $K$ representing its kernel.
In addition to $P$ we also need as input an array containing the outwards normals of its faces, as it is not always possible to determine the orientation of a face only from its vertices (for example with non-convex faces).
We require the face normals explicitly, and not simply a boolean indicating the faces orientation, because these points will be used to define the planes containing the faces of $P$.

We initialize $K$ with the axis aligned bounding box (AABB) of $P$, i.e. the box with the smallest volume within which all the vertices of $P$ lie, aligned with the axes of the coordinate system.
Then we recursively ``slice'' $K$ with a number of planes, generating a sequence of convex polyhedra $K_i$, $i=1,\dots,\#$\textit{P.faces}, such that $K_i\subseteq K_{i-1}$.
For each face $f$ of $P$ we define the plane $p$ containing its vertices and with normal vector given by the opposite of the face normal $N(f)$, that is to say, $p.n:=-N(f)$.
We consider the plane together with the direction indicated by $p.n$, which is equivalent to considering the half-space originating in $p$ and containing its normal vector.
Given this plane, in a \textit{Sign} array $S$ we store the classification of all the points in $K.verts$ according to their position with respect to $p$.
For improving the efficiency, we can set the sign of all points belonging to $f$ to zero without evaluating their position.
In general $p$ will separate $K$ into two polyhedra, and between those two we keep the one containing the vector $p.n$, which therefore points towards the interior of the element.
This operation is performed by the \textit{Polyhedron-Plane Intersection} algorithm detailed in Section \ref{subsec:polygon_plane}, which replaces $K$ with the new polyhedron.

The order in which we consider the faces is not relevant from a theoretical point of view, but turns out to have a huge impact on the performance.
For instance, if we imagine to compute the kernel of a ``ring'', which is obviously empty, visiting the faces in the order they are stored may take a very long time, especially if the tessellation of the object is fine.
This is because, generally, the faces of a tessellated model are numbered somehow coherently with their neighbors.
For this reason, we optionally propose to visit the faces in random order, or \emph{shuffle} $P.faces$, and to return an empty polyhedron if after a ``slice'' $K$ has less than three faces.
In this way, the empty kernel of a ring with thousands of faces could be detected in just three or four iterations.
When this command is turned on, we say the algorithm is run in \textit{shuffle} mode.

We point out that cutting a convex polyhedron with a plane will always generate two convex polyhedra, and since we start from the bounding box (which is convex), we are guaranteed that $K$ will always be a convex polyhedron.
No matter how weird the initial element $P$ is, from this point on we will only be dealing with convex polyhedra and convex faces.
Last, we could as well start with considering the polyhedron's convex hull instead, but it would be less efficient because the convex hull costs in general $O(n\log n)$ while the AABB is only $O(n)$, and we would still need to intersect the polyhedron with each of its faces.

\begin{algorithm}[htbp]
\caption{Polyhedron Kernel}
\label{alg:kernel}
\begin{algorithmic}[1]
\Require Polyhedron $P$, Points $N$ (faces normals);
\Ensure Polyhedron $K$
\State $K$ := AABB of $P$;
\State \textit{[optional]} shuffle $P.faces$;
\For{Face $f$ in $P.faces$}
    \State Plane $p$ := plane containing $f$ with normal $-N(f)$;
    \State Sign $S$ := orient3d$(p.p_1, p.p_2, p.p_3, K.verts)$;
    \State $K$ := Polyhedron-Plane Intersection($K$, $S$, $p$);
    \If{size($K.faces$) $<3$} return NULL;
    \EndIf
\EndFor
\State \Return $K$;
\end{algorithmic}
\end{algorithm}


\subsection{Polyhedron-Plane Intersection}
\label{subsec:polyhedron_plane}
With the second algorithm we want to intersect a polyhedron $P$ with a plane $p$, given in $S$ the position of the vertices of $P$ with respect to $p$.
The intersection will in general determine two polyhedra, and between these two we are interested in the one containing the normal vector of $p$ (conventionally called the one ``above'' the plane and indicated with $A$).
This algorithm is inspired from \cite{ahn2008geometric}, where the authors define an algorithm for the intersection of a convex polyhedron with an half-space.

\begin{figure}[htbp]
\centering
\begin{tabular}{cc}
\includegraphics[width=.46\linewidth]{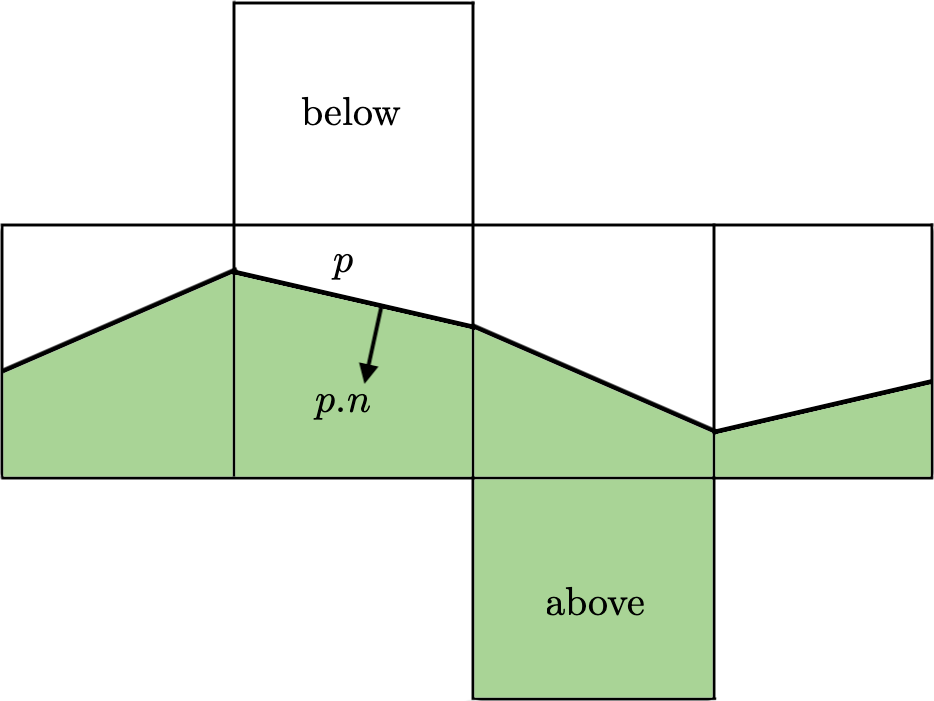} &
\includegraphics[width=.46\linewidth]{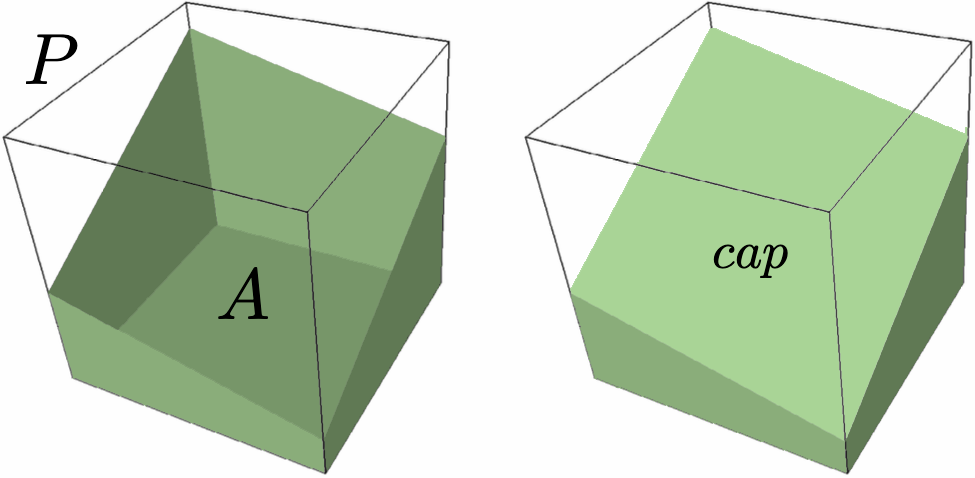}\\
(a) & (b)
\end{tabular}
\caption{Intersection of a polyhedron with a plane: (a) clipping and (b) capping of a cube.}
\label{fig:polyhedron-plane}
\end{figure}

The first part of Algorithm~\ref{alg:polyhedron-plane} is called the ``clipping'' part (recalling the terminology from \cite{ahn2008geometric}) and consists in clipping each face of $P$ with the plane $p$, see Fig.~\ref{fig:polyhedron-plane}(a).
It corresponds to the \textit{for} loop: we iterate on \textit{P.faces}, each time extracting from $S$ the labels $f_s$ of the vertices $f_v$ of the current face and using the \textit{classify} function.
Faces classified as BELOW are discarded, ABOVE faces are added to $A$ together with their vertices, and INTER faces are split by the \textit{Polygon-Plane Intersection} algorithm.
While we visit every face only once, the same does not hold for vertices, therefore we check if a vertex is already in $A.verts$ before adding it.

This simple idea of checking in advance the faces classification resolves several implementation issues and in some cases significantly improves the efficiency of the algorithm.
By doing this, we make sure that only the faces properly intersected by the plane are passed to Algorithm~\ref{alg:polygon-plane}, so that we do not need to implement all the particular cases of intersections in a single point or along an edge or of faces contained in the plane.
In addition, for every face not passed to Algorithm~\ref{alg:polygon-plane} we have an efficiency improvement, and this happens frequently in models with many coplanar faces like the ones considered in Section~\ref{subsec:refinements}.

If, at the end of this step, $A$ contains at least three INTER points, given that $A$ and all its faces are convex, these vertices will define a ``cap'' face of $A$ completely contained in $p$, see Fig.~\ref{fig:polyhedron-plane}(b).
We can optimize the algorithm by storing in a Sign array the classification of the vertices in $A$, updating it with the sign of every vertex added in the switch loop.
Note that in this case we do not need to use \textit{orient3d}: we already know the sign of the old vertices and the new vertices will obviously be of type INTER.
In our data structure, the vertices of the faces are ordered counter-clockwise (CCW): in order to sort the points contained in \textit{capV} we project them onto a plane, drop one coordinate and apply the algorithm proposed in \cite{baeldung} for 2D points.
Note that if the cap face was not convex it would make no sense to order its vertices, but the intersection between a plane and a convex polyhedron will always generate convex faces.
Last, we need to check that this new face is not already present in $A$: for example if $p$ was tangent to $P$ along a face, this face could be added to $A$ both as an ABOVE face and as a cap face.
If this is not the case, we add \textit{capF} to $A.faces$ but we do not need to add any vertex from \textit{capV}, as we can assume they are all already present in $A.verts$.

\begin{algorithm}[htbp]
\caption{Polyhedron-Plane Intersection}
\label{alg:polyhedron-plane}
\begin{algorithmic}[1]
\Require Polyhedron $P$, Sign $S$, Plane $p$
\Ensure Polyhedron $A$
\For{Face $f$ in $P.faces$}
    \State Points $f_v:=$ vertices in $P.verts$ relative to $f$;
    \State Sign $f_s$ := $S_{|f_v}$
    \Switch{classify$(f_s)$}
    \Case{BELOW} break;
    \EndCase
    \Case{ABOVE}
        \State $A.verts\leftarrow$ $f_v$, $A.faces\leftarrow f$;
    \EndCase
    \Case{INTER}
      \State (\textit{V},\textit{F}):=Polygon-Plane Intersection$(f_v,f,f_s,p)$;
        \State $A.verts\leftarrow$ \textit{V},
        $A.faces\leftarrow$ \textit{F};
    \EndCase
    \EndSwitch
\EndFor
\State Points \textit{capV} := vertices in $A.verts$ which are INTER;
\If{size(\textit{capV}) $<3$} return $A$;
\EndIf
\State Face \textit{capF} := indices of \textit{capV} vertices ordered CCW;
\If{\textit{capF} $\notin$ $A.faces$}
    $A.faces\leftarrow$ \textit{capF};
\EndIf
\State \Return $A$;
\end{algorithmic}
\end{algorithm}


\subsection{Polygon-Plane Intersection}
\label{subsec:polygon_plane}
\begin{figure}[htbp]
\centering
\begin{tabular}{cc}
\includegraphics[width=.35\linewidth]{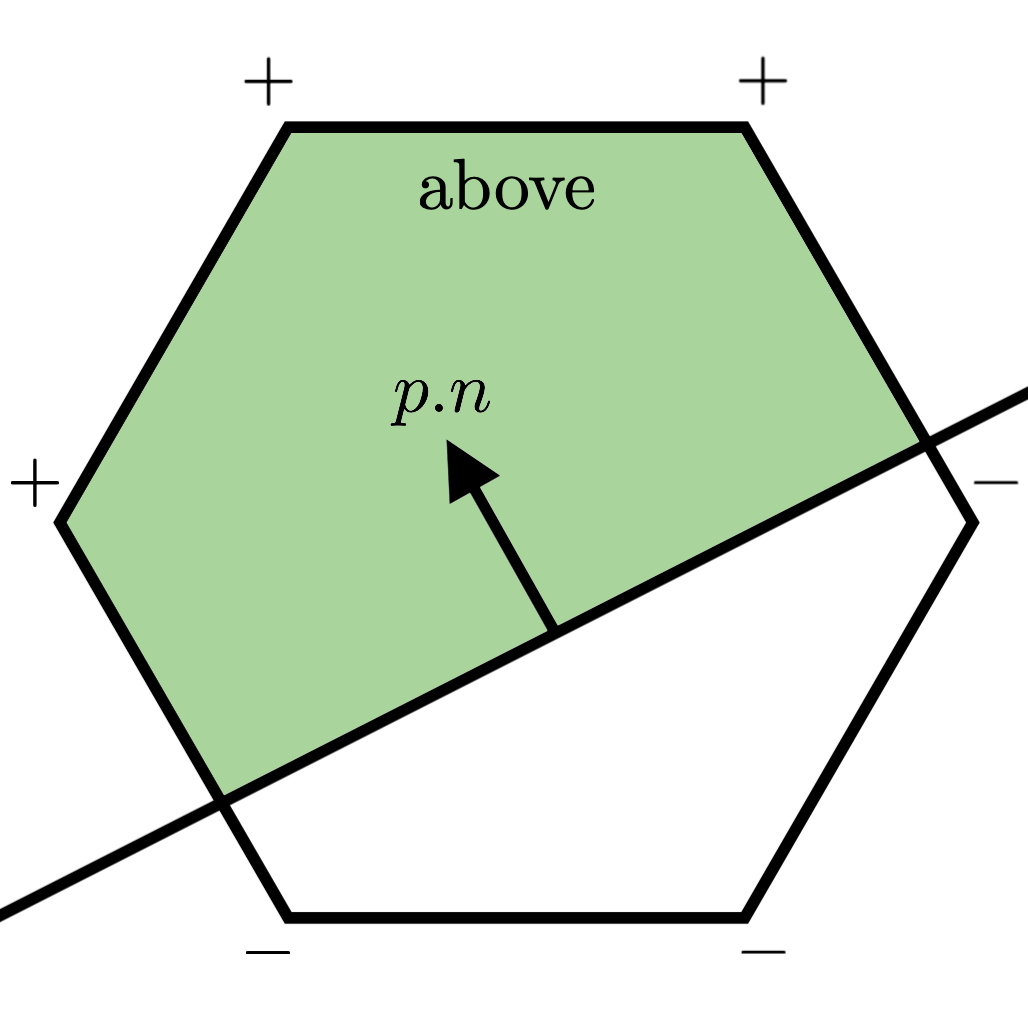} &
\includegraphics[width=.35\linewidth]{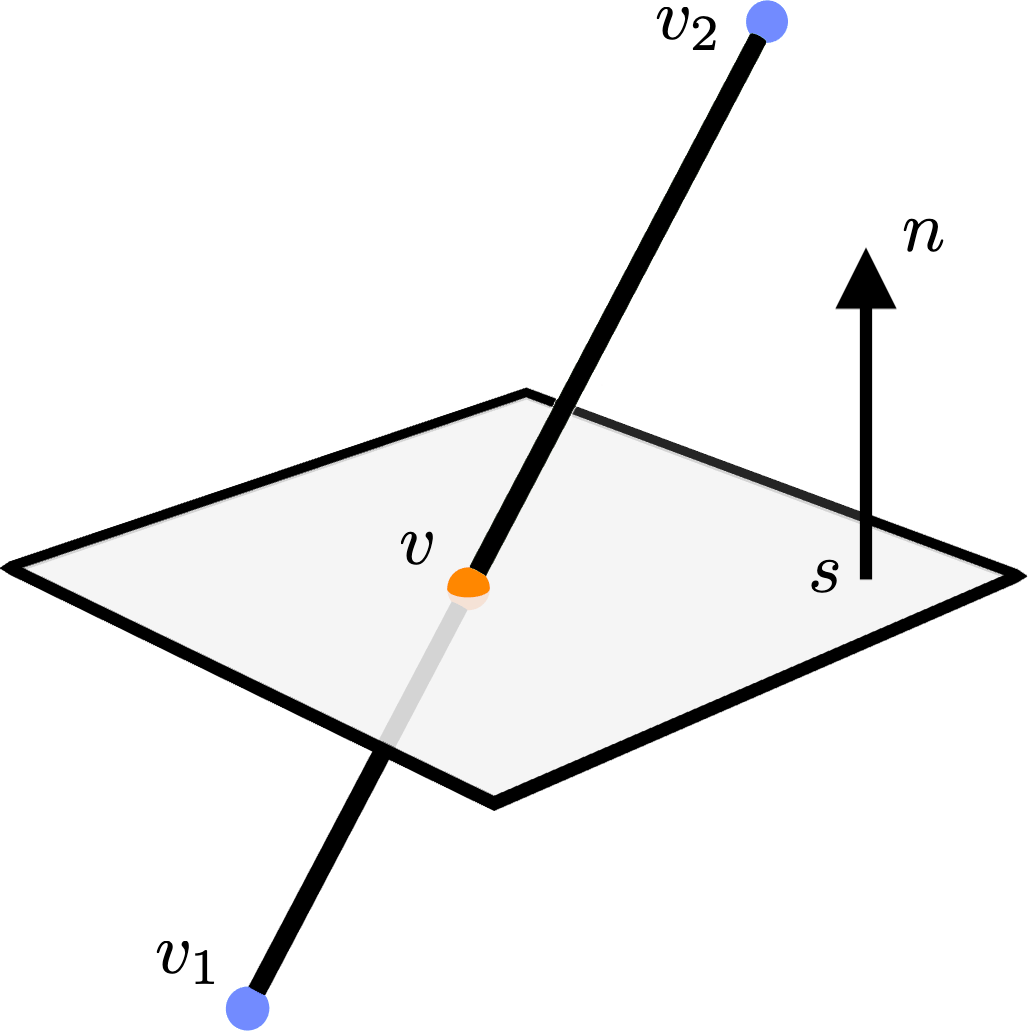}\\
(a) & (b)
\end{tabular}
\caption{(a) Intersection between a polygon and a plane, with the above part coloured in green. (b) Intersection between a line and a plane.}
\label{fig:polygon-line-plane}
\end{figure}

Algorithm~\ref{alg:polygon-plane} describes the intersection of a polygon (representing a face of the polyhedron), defined by an array of 3D points \textit{polyV} and an array of indices \textit{polyF}, with a plane $p$.
As before, we also require as input an array \textit{polyS} containing the position of the vertices of \textit{polyV} with respect to $p$.
In analogy to Algorithm~\ref{alg:polyhedron-plane}, the intersection will in general determine two polygons and we are only interested in the one above the plane, see Fig.~\ref{fig:polygon-line-plane}(a), defined by points \textit{aboveV} and indexes \textit{aboveF}.
We generically say that a vertex $v$ is added to \textit{above} meaning that $v$ is added to \textit{aboveV} and its index $id_v$ is added to \textit{aboveF}.

This time we iterate on the edges of \textit{polyF}, extract the signs $s_1, s_2$ of the edge endpoints and switch between the three possible classifications of the edge.
In order to avoid duplicates, for each couple of consecutive points $v_1, v_2$, we only accept to add to \textit{above} the second point $v_2$ or the intersection point $v$, but never $v_1$.
We are here taking advantage of the fact that all faces are oriented coherently.

If the edge is of type BELOW we ignore it unless $v_2$ is INTER (i.e. it lies exactly on the plane), in which case we add it to \textit{above}.
In case of ABOVE edges we add $v_2$ to \textit{above}.
For edges of type INTER we perform the \textit{Line-Plane Intersection} algorithm and find a new point $v$.
Its index $id_v$ will be equal to the maximum value in \textit{polyF} plus one, just to make sure that we are not using the index of an existing point.
We always add $v$ to \textit{above}, and if $v_1$ is BELOW we also add $v_2$.
As already noted in Section~\ref{subsec:polyhedron_plane}, treating separately the weak intersections (the BELOW and ABOVE cases) makes the code simpler and more efficient.

\begin{algorithm}[htbp]
\caption{Polygon-Plane Intersection}
\label{alg:polygon-plane}
\begin{algorithmic}[1]
\Require Points \textit{polyV}, Face \textit{polyF}, Sign \textit{polyS}, Plane $p$.
\Ensure Points \textit{aboveV}, Face \textit{aboveF}.
\For{$i=1$ : size(\textit{polyF})}
     \State $id_1:=$ \textit{polyF}$(i)$, 
     $id_2:=$ \textit{polyF}$(i+1)$;
     \State $v_1:=$ \textit{polyV}$(id_1)$, 
     $v_2:=$ \textit{polyV}$(id_2)$;
     \State $s_1:=$ \textit{polyS}$(id_1)$, 
     $s_2:=$ \textit{polyS}$(id_2)$;
     \Switch{classify($s_1,s_2$)}
     \Case{BELOW}
         \If{$v_2$ is INTER}
            \State \textit{aboveV} $\leftarrow v_2$, \textit{aboveF} $\leftarrow id_2$;
        \EndIf
    \EndCase
    \Case{ABOVE}
        \State \textit{aboveV} $\leftarrow v_2$, \textit{aboveF} $\leftarrow id_2$;
    \EndCase
    \Case{INTER}
        \State $v:=$ Line-Plane Intersection$(v_1,v_2,p)$;
        \State $id_v:=$ max(\textit{polyF})+1;
        \State \textit{aboveV} $\leftarrow v$, \textit{aboveF} $\leftarrow id_v$;
        \If{$v_1$ is BELOW}
            \State \textit{aboveV} $\leftarrow v_2$, \textit{aboveF} $\leftarrow id_2$;
        \EndIf
    \EndCase
     \EndSwitch
\EndFor
\State \Return \textit{aboveV}, \textit{aboveF};
\end{algorithmic}
\end{algorithm}


\subsection{Line-Plane Intersection}
\label{subsec:line_plane}
This last algorithm computes the intersection point between a line, given as a couple of vertices, and a plane.
It is a very simple and well known procedure, and we report it here only for completeness.

The intersection vertex $v$ is defined as the linear combination of vertices $v_1$ and $v_2$, with a coefficient $t$ which may also fall outside the standard range $[0,1]$.
The coefficient $t$ is found as the negative ratio between two scalar products involving the plane normal $n$ and a generic other point on the plane $s$, other than $v_1$ and $v_2$ see Fig~\ref{fig:polygon-line-plane}(b).
The normal is $p.n$, and for the point $s$ we can use one of the three points on the plane $p.p_1, p.p_2, p.p_3$.
If the denominator $D$ vanishes, it means either that the line is contained in the plane (if $N=0$ as well) or that the line does not intersect the plane.
We treat these exceptions as errors because in Algorithm~\ref{alg:polygon-plane} we only call this algorithm after checking that the edge $(v_1,v_2)$ properly intersects the plane $p$.

\begin{algorithm}[htbp]
\caption{Line-Plane Intersection}
\label{alg:line-plane}
\begin{algorithmic}[1]
\Require vertices $v_1,v_2$, Plane $p$.
\Ensure vertex $v$.
\State $N:=(p.n)\cdot(v_1-p.p_1)$;
\State $D:=(p.n)\cdot(v_2-v_1)$;
\Assert{$D!=0$}
\State $t:=-N/D$;
\State \Return $v:=v_1+t\ (v_2-v_1)$;
\end{algorithmic}
\end{algorithm}


\subsection{Computational complexity}
\label{subsec:computational_complexity}
Making advantage of the modular organisation of our algorithms, we can estimate separately the computational cost of each algorithm and then include them into a single formula.

Let us consider the case of an input polyhedron $P$ with $n_v$ vertices and $n_f$ faces.
In Algorithm~\ref{alg:kernel} we start by computing the polyhedron's AABB, which is $O(n_v)$, then for each face we estimate the position of the vertices of $K$ with respect to a plane with \textit{orient3d}.
Unfortunately, we have to re-compute the signs of all the $n_{kv}$ vertices of $K$ at every iteration, as from one step to the other the plane changes: this means $O(n_{kv} n_f)$ operations.
At step zero we have $n_{kv}=8$, being $K$ the bounding box; then in the worst case, that is if the object is convex, it can grow up to $n_v$ because $K$ coincides with $P$.
If the object is not convex, $n_{kv}$ can remain significantly lower than $n_v$, which translates in a much lower computational cost.
In both cases, the difference between $n_{kv}$ and $n_v$ is also related to the number of coplanar faces of the model, which get agglomerated into a single face in the kernel with a consequent reduction of the number of vertices.
If the kernel is empty, whenever we end up with less than three faces the algorithm stops, therefore we can replace $n_f$ with the number of iterations needed to detect the non star-shapedness.
This number cannot be estimated precisely, but we can drastically reduce it with the shuffle mode.

The good news is that once that we have the sign of all the vertices with respect to all the planes we are essentially done: the computational costs of Algorithms~\ref{alg:polyhedron-plane} and~\ref{alg:polygon-plane} is all about visiting arrays and copying parts of them into other arrays, and Algorithm~\ref{alg:line-plane} only consists of 4 operations.
We point out that navigating and duplicating arrays has a negligible cost in this scenario as the faces of a generic polyhedron hardly contain more than 10 vertices.
The only relevant operation in Algorithm~\ref{alg:polyhedron-plane} is the sorting of the vertices of the cap face, which does not always exist.
This is done with a QuickSort routine which is on average $O(n\log n)$ for a cap face with $n$ vertices.
Since $n$ is much smaller than $n_{kv}$ and the number of cap faces is always smaller than $n_f$, this cost is negligible compared to $O(n_{kv} n_f)$.

In summary, we can set an upper bound to the computational cost of our algorithm at $O(n_v n_f)$, but both $n_v$ and $n_f$ can significantly decrease if the model is not convex or has coplanar faces, and even more if it is not star-shaped.
In the next section we will show how in practice, on small polyhedra or on objects with many co-planar faces, the geometric method works in a computational time which is smaller than the one of the algebraic approach, and it is still competitive on many examples, even complex ones.

\section{Tests and discussions}
\label{sec:tests}
We test our method in different settings, comparing its performance to the results obtained using our implementation of the algebraic method in CGAL. The comparison of the current method, including the shuffle version, and its previous version in \cite{SorgenteKernel} is presented in Section \ref{subsec:comparison}. 

Experiments have been performed on a MacBook Pro equipped with a 2,3 GHz Intel Core i5 processor with four CPUs and 16GB of RAM.
Source code is written in C++ and it is accessible at \url{https://github.com/TommasoSorgente/polyhedron_kernel}, together with all datasets.

In the following subsections we will present some plots and tables: we point out here some remarks on the notation adopted.
Regarding plots, we colour the CGAL computational time in blue and our computational time in red, both in logarithmic scale.
On the $x-$axis, depending from the context, we have the number of elements in the mesh or the number of vertices of the single model.
In the several tables presented we first report the number of elements or vertices of the mesh.
Since all the considered objects have genus zero and their surface is purely triangular, Euler's formula states that the number of faces is approximately equal to twice the number of vertices.
Therefore we only indicate the number of vertices, but the number of faces is easily computable.
Then the computational times (in seconds) are shown, and the ratio between CGAL time and ours.
Note that ratios are computed from the original time values while in the tables we indicate truncated times, therefore they do not exactly correspond to the division between the values in the previous columns.


\subsection{Polyhedral meshes}
\label{subsec:meshes}
\begin{figure*}[htbp]
\centering
\includegraphics[width=\linewidth]{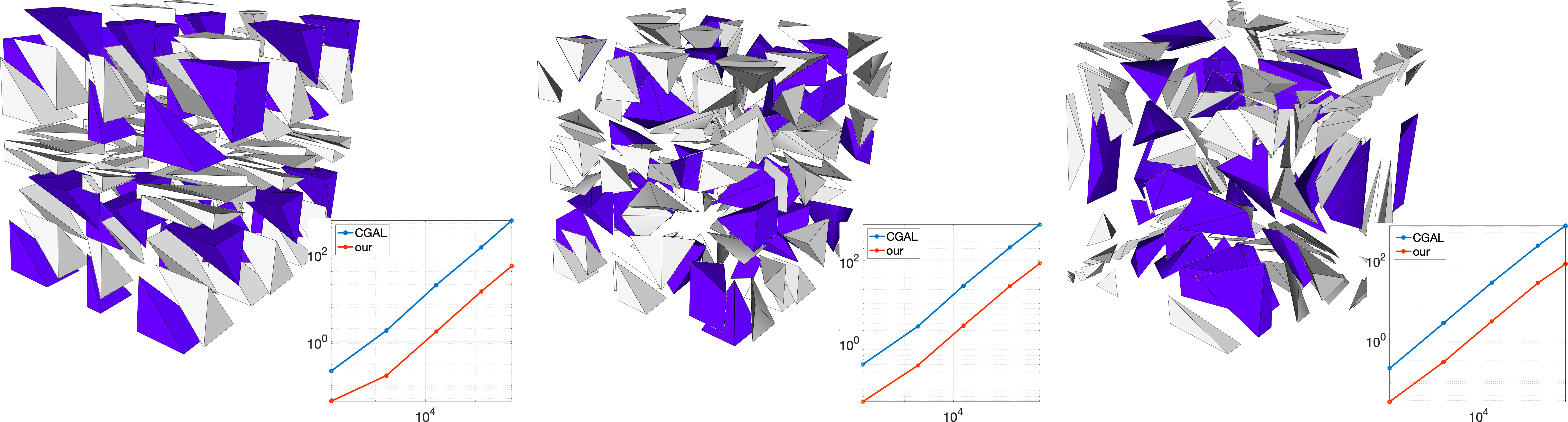}
\caption{Polyhedral meshes and time plots from datasets \textit{poly-parallel}, \textit{poly-poisson} and \textit{poly-random}, with non-tetrahedral elements highlighted in blue.}
\label{fig:meshes}
\end{figure*}
First, we test our algorithm in the setting it was developed for, i.e. the computation of the kernels of elements in a 3D tessellation.
To do so, we used the datasets from \cite{sorgente2021polyhedral}, available for download at \url{https://github.com/TommasoSorgente/vem-indicator-3D-dataset}.
The meshes contained in these datasets are typical examples of tessellations which can be found in numerical analysis for the approximation of a PDE.
We focus our attention on the polyhedral datasets: \textit{poly-parallel}, \textit{poly-poisson} and \textit{poly-random}.
Each of them contains five tessellations of the unit cube with decreasing mesh size, from 100 to 100K vertices.
The resulting meshes contain between 100 and 600K elements, most of which are tetrahedra but $20\%$ of them are generic polyhedra (in blue in Fig.~\ref{fig:meshes}), obtained by the union of two tetrahedra.

\begin{figure}[htbp]
\centering
\includegraphics[width=\linewidth]{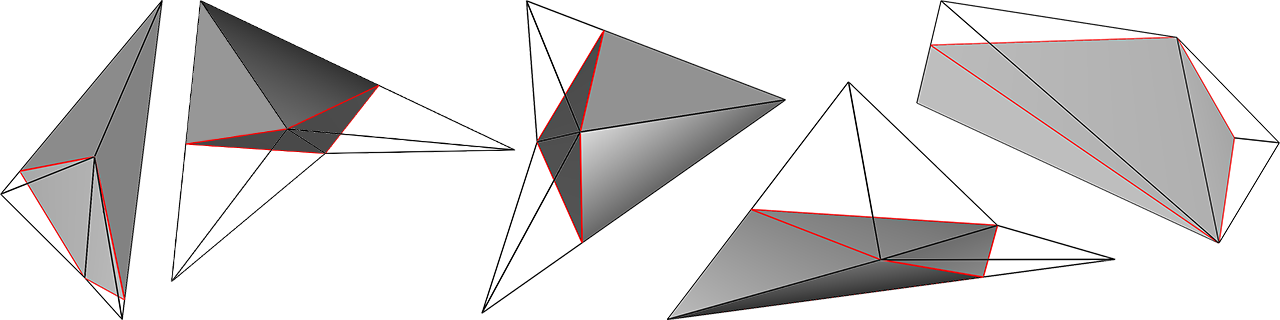}
\caption{Examples of non-convex elements found in polyhedral meshes from Section~\ref{subsec:meshes} and relative kernels.}
\label{fig:elements}
\end{figure}

Non-tetrahedral elements are generated by the agglomeration of two tetrahedral elements, therefore they may also be non convex, see Fig.~\ref{fig:elements}.

\begin{table}[htbp]
\caption{Computational times for polyhedral meshes.}
\label{table:time:mesh}
\centering
\begin{tabular}{ccccc}
\hline\noalign{\smallskip}
dataset & $\#$elements & our & CGAL & ratio\\
\noalign{\smallskip}\hline\noalign{\smallskip}
\textit{poly-parallel} 
& 130    & 0.04  & 0.21   & 4.89 \\
& 1647   & 0.17  & 1.79   & 10.69 \\
& 16200  & 1.68  & 19.4   & 11.55 \\
& 129600 & 13.94 & 142.36 & 10.21 \\
& 530842 & 53.47 & 588.43 & 11 \\
\noalign{\smallskip}\hline
\textit{poly-poisson} 
& 140    & 0.04  & 0.3    & 8.05 \\
& 1876   & 0.29  & 2.54   & 8.91 \\
& 16188  & 2.64  & 24.79  & 9.38 \\
& 146283 & 24.24 & 212.23 & 8.75 \\
& 601393 & 86.77 & 770.66 & 8.88 \\
\noalign{\smallskip}\hline
\textit{poly-random} 
& 147    & 0.03  & 0.19   & 6.8 \\
& 1883   & 0.28  & 2.62   & 9.41 \\
& 18289  & 2.91  & 27.11  & 9.33 \\
& 161512 & 26.51 & 228.08 & 8.6 \\
& 598699 & 80.15 & 735.55 & 9.18 \\
\noalign{\smallskip}\hline
\end{tabular}
\end{table}

In Table~\ref{table:time:mesh} we report, for each mesh of each dataset, the number of elements, the computational times for both methods and the ratio between the CGAL time and ours. 
Moreover, at the bottom of Fig.~\ref{fig:meshes} we plot the times against the number of elements in the mesh.
It is visible how both methods scale linearly with respect to the number of elements, since the kernel of the elements is computed separately and independently for each element.
Our method performs 8 to 11 times faster than CGAL, which approximately means one order of magnitude.
As the elements of these meshes have either 4 faces if they are tetrahedra or 6 faces if they are the union of two tetrahedra, computing their kernel in a geometrical way results much faster than solving a linear problem.
In this case we did not use the shuffle mode, as the number of faces was so small that the visiting order resulted not relevant.


\subsection{Refinements}
\label{subsec:refinements}
As a second setting for our tests, instead of increasing the number of elements we wanted to measure the asymptotic behaviour of the methods as the number of faces and vertices of a single element explodes.
We selected two polyhedra from the dataset \textit{Thingi10K} \cite{zhou2016thingi10k}: the so-called \textit{spiral} (ThingiID: 60246) and \textit{vase} (ThingiID: 85580).
These models are given in the form of a surface triangular mesh but we treat them as single volumetric cell, analyzing the performance of both algorithms as we refine them.
In Table~\ref{table:time:refinements} we report the computational times and the ratio for each refinement.

\begin{figure}[th]
\centering
    \includegraphics[width=.8\linewidth]{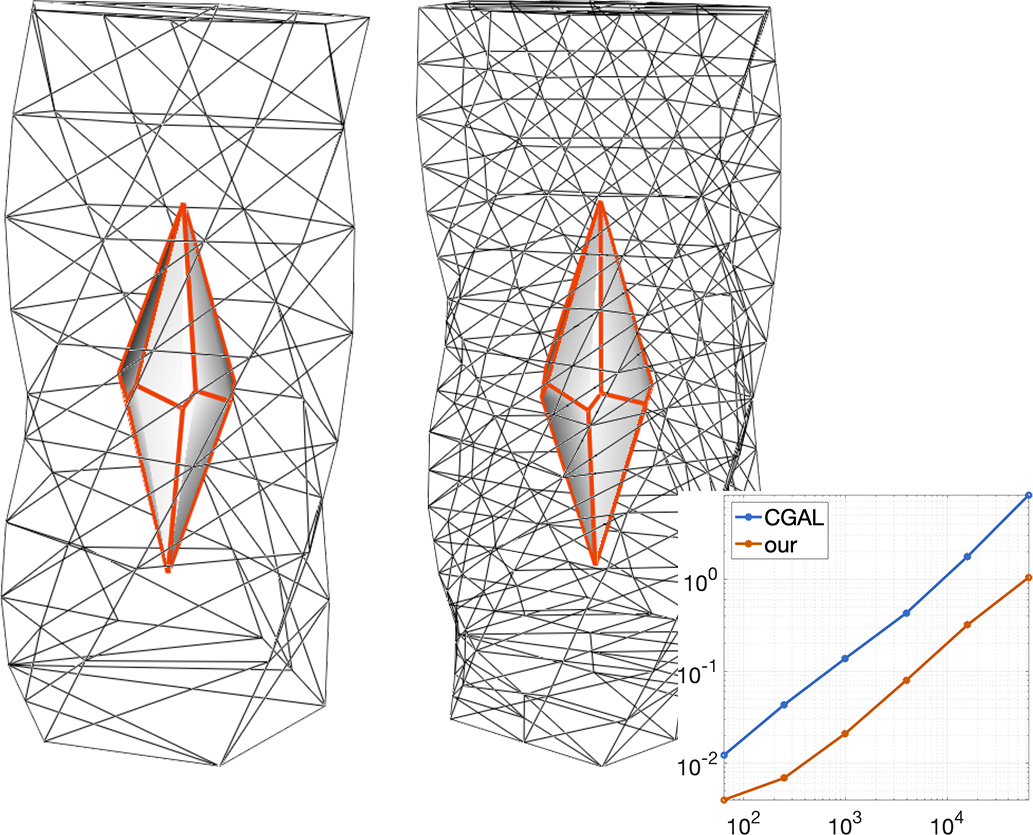}
\caption{Original \textit{spiral} model and its first refinement, with identical kernels.}
\label{fig:spiral}
\end{figure}

The \textit{spiral} model is refined through a midpoint strategy: each face is subdivided by connecting its barycenter to its other vertices.
As a consequence, the planes induced by its faces remain the same and the kernels of the refined models are all equal (Fig.~\ref{fig:spiral}).
On this example our method performs on average 5.77 times better that the algebraic method (see Table~\ref{table:time:refinements}), and the computational time scales with a constant rate (see the plot in Fig.~\ref{fig:spiral}).
Our implementation takes advantage of the fact that Algorithm~\ref{alg:polyhedron-plane} recognises the several coplanar faces and always performs Algorithm~\ref{alg:polygon-plane} the same number of times, independently of the number of faces.

\begin{figure}[htbp]
\centering
    \includegraphics[width=\linewidth]{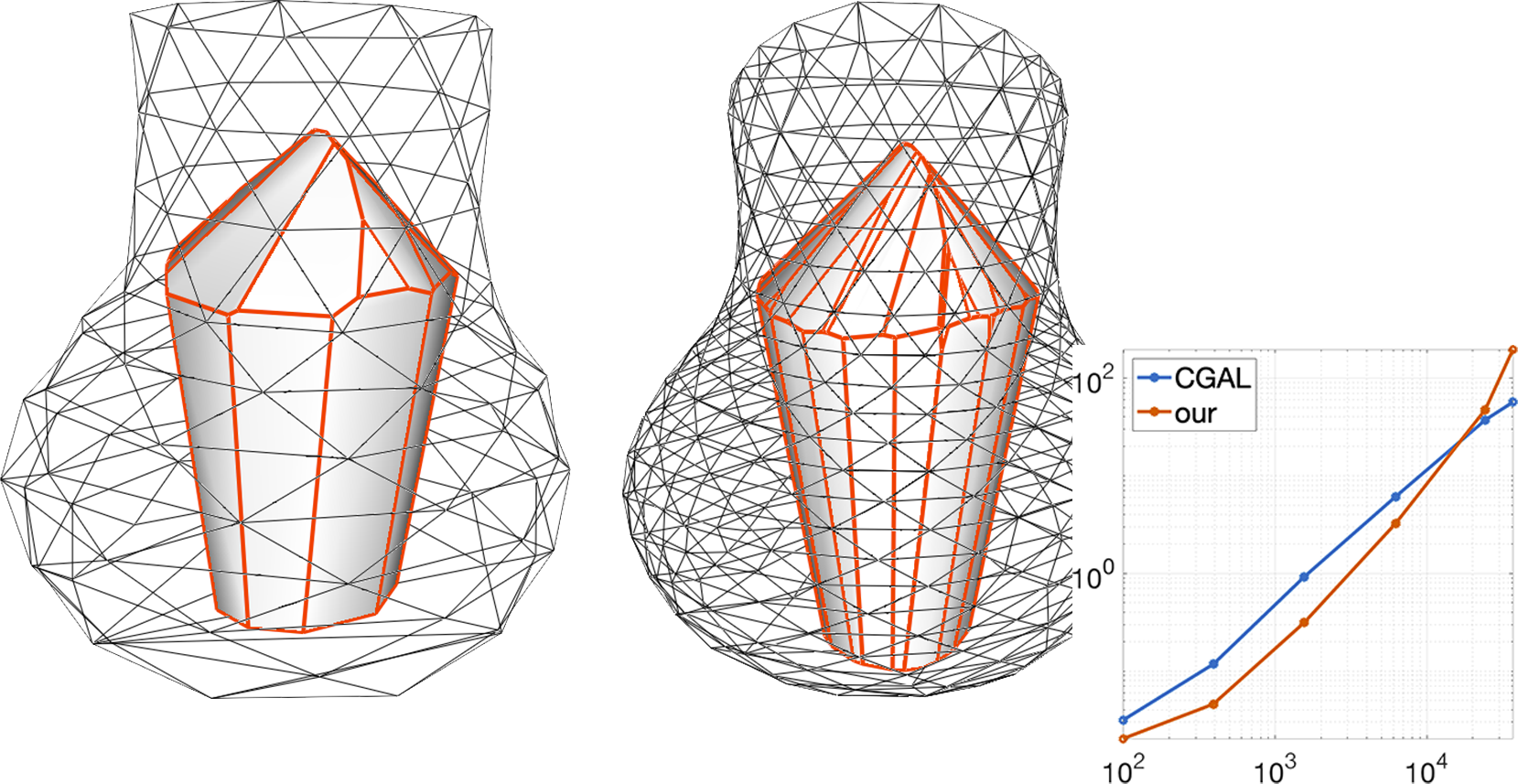}
\caption{Original \textit{vase} model and its first refinement: small perturbations in the faces lead to slightly different kernels.}
\label{fig:vase}
\end{figure}

The \textit{vase} model is more complex, as it presents a curved surface which generates a lot of different planes defining the kernel.
Moreover, we refined this model using the Loop's algorithm and this generated faces lying on completely new planes.
This explains the difference between the two kernels in Fig.~\ref{fig:vase}: the general shape is similar but the more faces we add to our model the more faces we find on the resulting kernel.
Our geometric method improves the performance of the algebraic one by a factor around 2 in the first refinements, but in the last two meshes the complexity increases drastically and CGAL results faster (see Table.~\ref{table:time:refinements}).
In this case, an efficient treatment of the faces is not sufficient to hide the quadratic nature of the geometric approach.
Even the shuffle mode did not particularly improve the performance, being the object star-shaped.

\begin{table}[htbp]
\caption{Computational times for the \textit{spiral} and \textit{vase} refinements.}
\label{table:time:refinements}
\centering
\begin{tabular}{ccccc}
\hline\noalign{\smallskip}
mesh & $\#$vertices & our & CGAL & ratio \\
\noalign{\smallskip}\hline\noalign{\smallskip}
\textit{spiral} 
& 64 & 0.004 & 0.01 & 3.07 \\
& 250 & 0.007 & 0.04 & 6.27 \\
& 994 & 0.02 & 0.14 & 6.56 \\
& 3970 & 0.08 & 0.43 & 5.38 \\
& 15874 & 0.32 & 1.77 & 5.54 \\
& 63490 & 1.05 & 8.24 & 7.86 \\
\hline
\textit{vase} 
& 99 & 0.02 & 0.03 & 1.55 \\
& 390 & 0.04 & 0.12 & 2.56 \\
& 1554 & 0.31 & 0.92 & 2.94 \\
& 6210 & 3.24 & 6.1 & 1.88 \\
& 24261 & 47.49 & 37.24 & 0.78 \\
& 36988 & 196.7 & 56.75 & 0.29 \\
\noalign{\smallskip}\hline
\end{tabular}
\end{table}

\subsection{Complex models}
\label{subsec:complex_models}
Last, we try to compute the kernel of some more complex models, taken again from the dataset \textit{Thingi10K} and treated as single volumetric cells.
Even if our method is designed for dealing with polyhedra of relatively small size, as we already saw in Section~\ref{subsec:refinements} our algorithms are still able to compute the kernel of objects with thousands of vertices and faces.
We filtered the \textit{Thingi10K} dataset selecting only ``meaningful'' models: objects with one connected component, genus zero, Euler characteristic greater than zero, closed, not degenerate and of size smaller than 1MB.
Note that, even applying these filters, the majority of the models are not star-shaped, i.e. with empty kernel.
We discarded a few models for computational and stability reasons, for instance because one of the two algorithms failed to process them.
The final collection, which we will call the \textit{Thingi dataset}, contains exactly 1806 distinct volumetric models.

\begin{figure*}[htbp]
\centering
    \includegraphics[width=.9\linewidth]{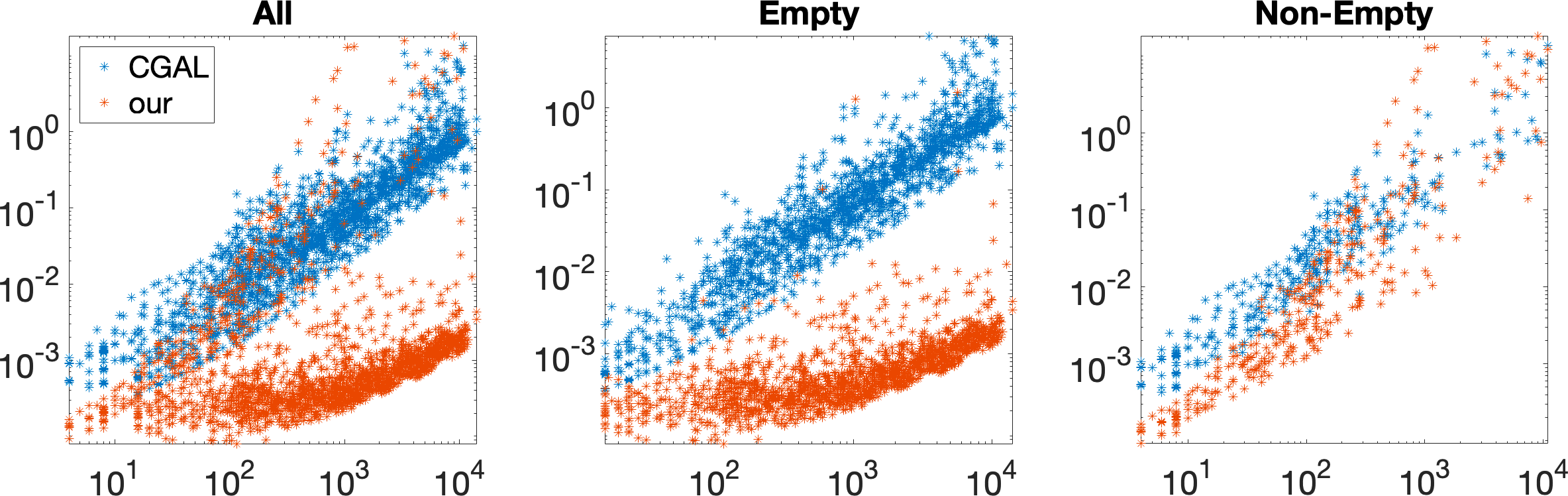} 
\caption{Thingi dataset times. From left to right: all Thingi dataset, models with empty kernel, models with non-empty kernel.}
\label{fig:time:thingi}
\end{figure*}

In Fig.~\ref{fig:time:thingi} we show the times distribution for the whole Thingi dataset, with a particular focus on the difference between models with empty kernel and models with non-empty kernel.
Globally, the overall cost of computing all kernels is 173 seconds with our method against 518 seconds with CGAL, for an improvement of 3 times.
When the kernel is empty our algorithm is always faster than CGAL: the main reason for this is the usage of the shuffle mode, which makes it extremely cheap to recognise non star-shaped elements.
When the model is star-shaped the distinction between the two methods is not so clear anymore, as the results mainly depend on the shape and size of the object.

\begin{figure*}[ht!]
\centering
\begin{tabular}{cccc}
\includegraphics[width=.2\linewidth]{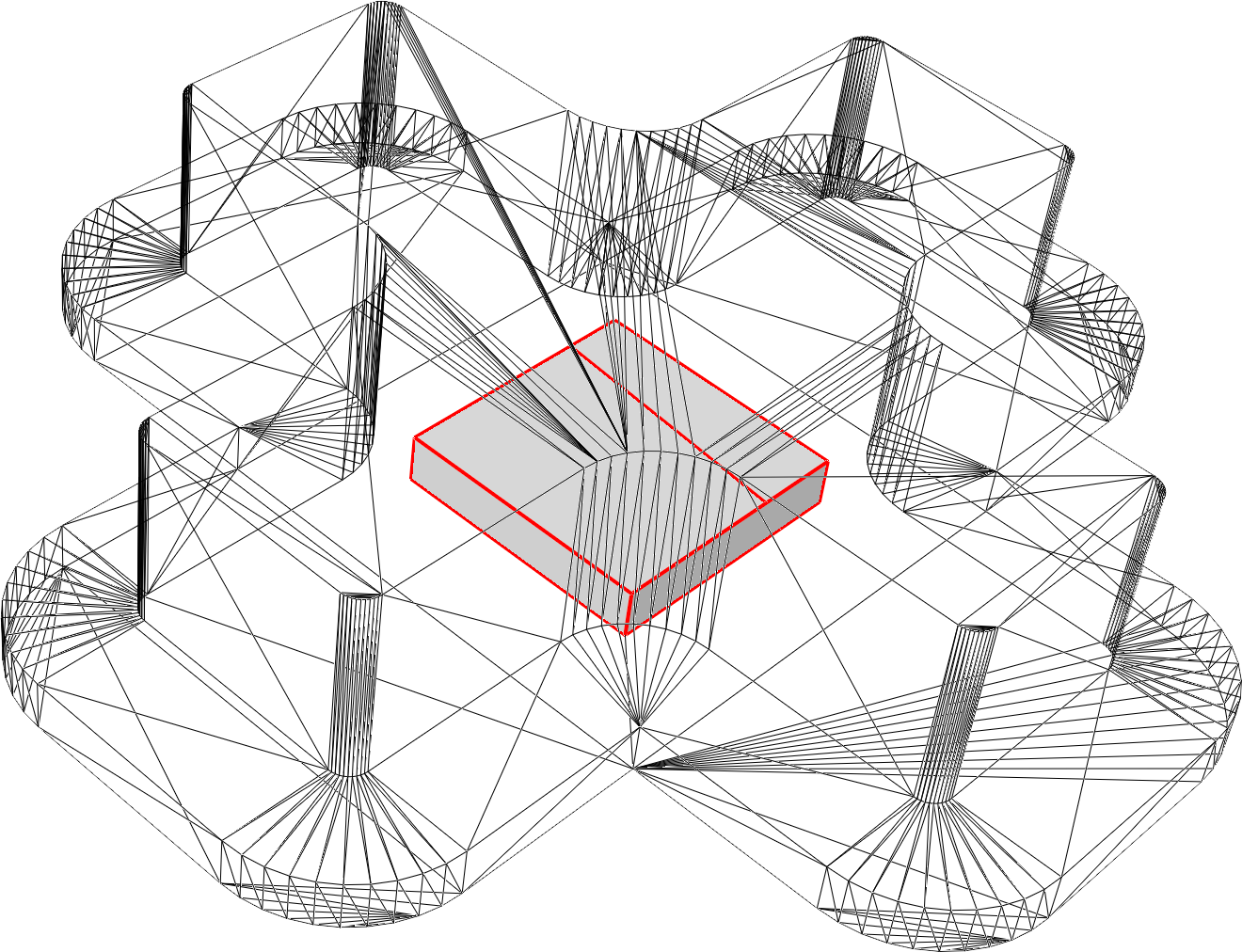} &
\includegraphics[width=.2\linewidth]{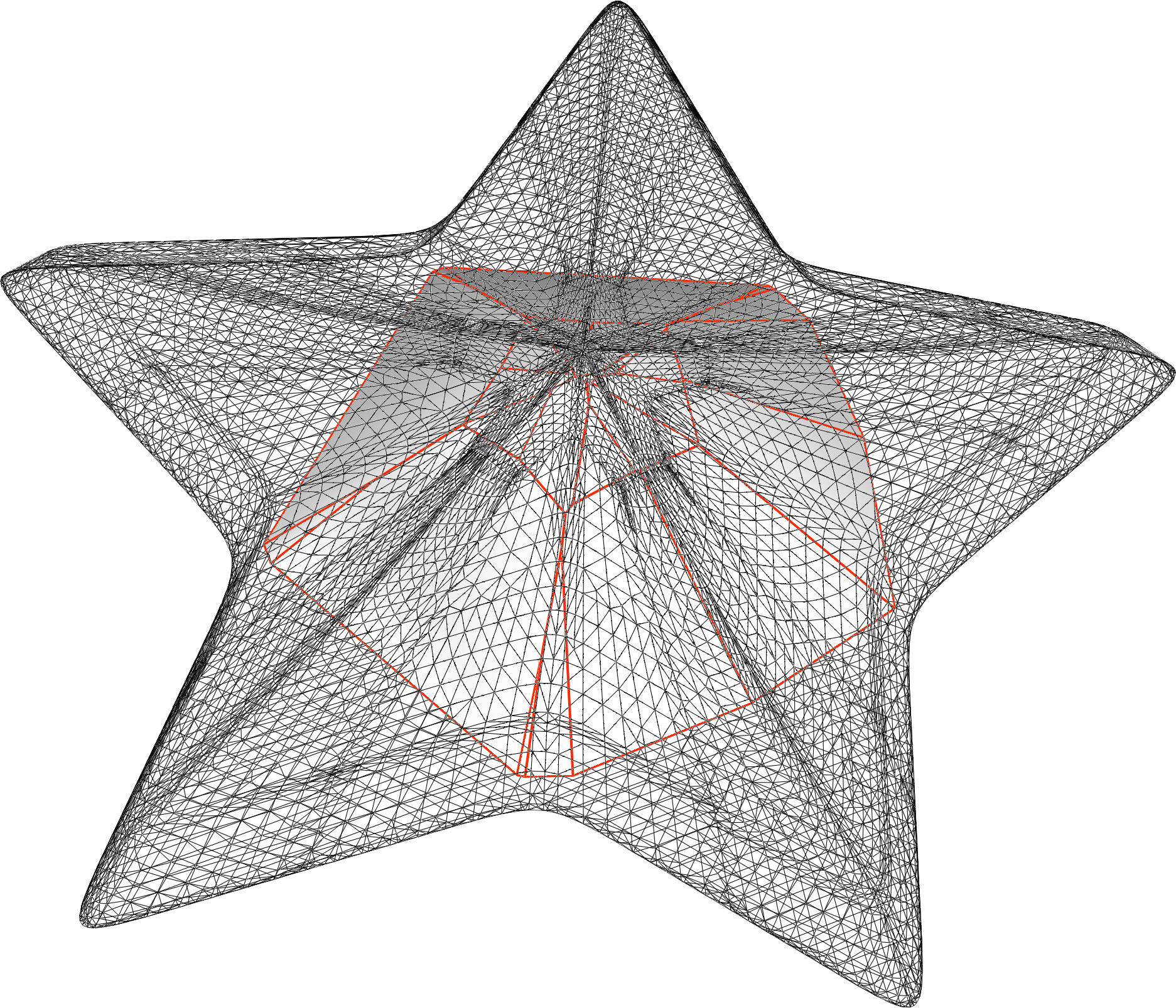} &
\includegraphics[width=.2\linewidth]{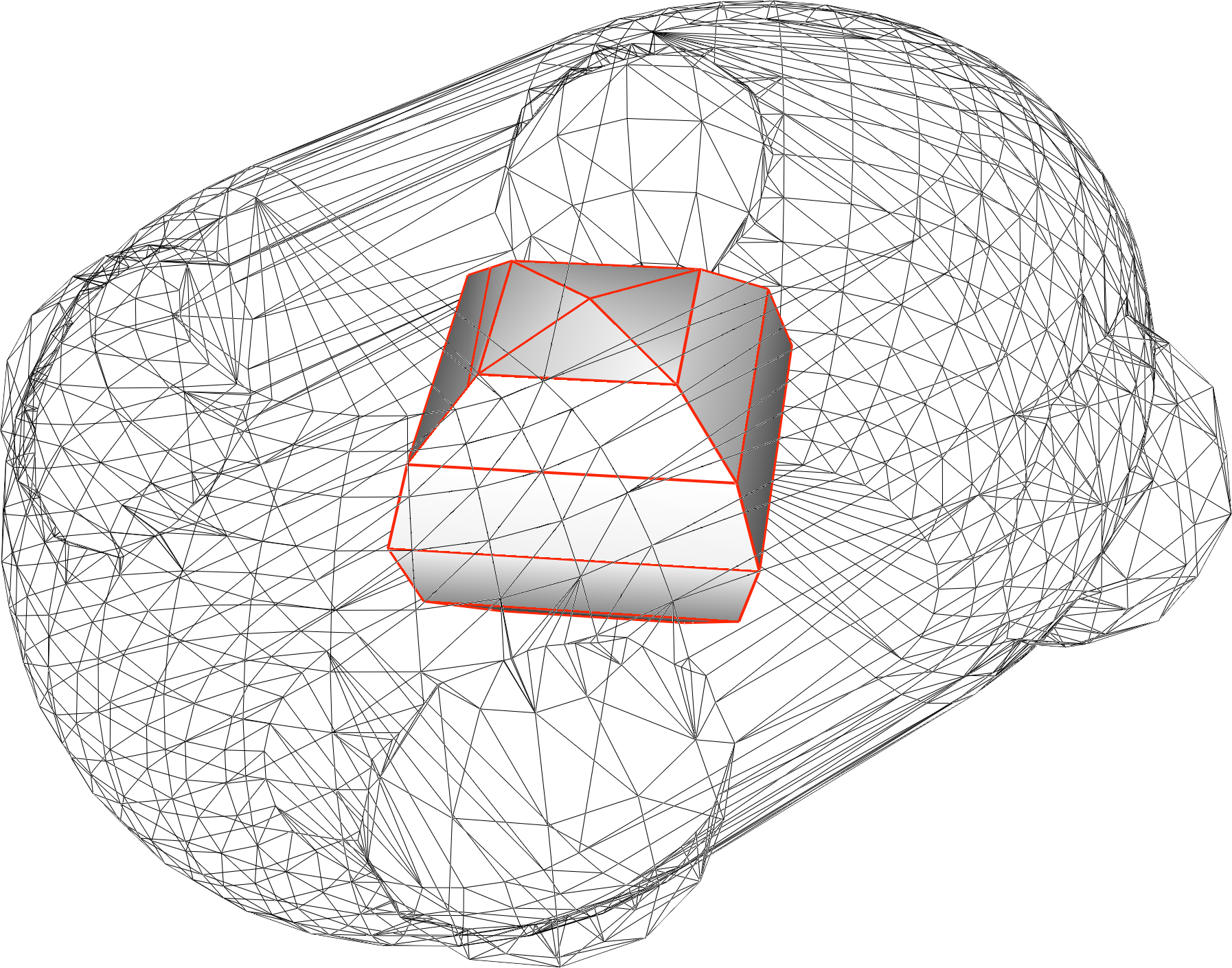} &
\includegraphics[width=.2\linewidth]{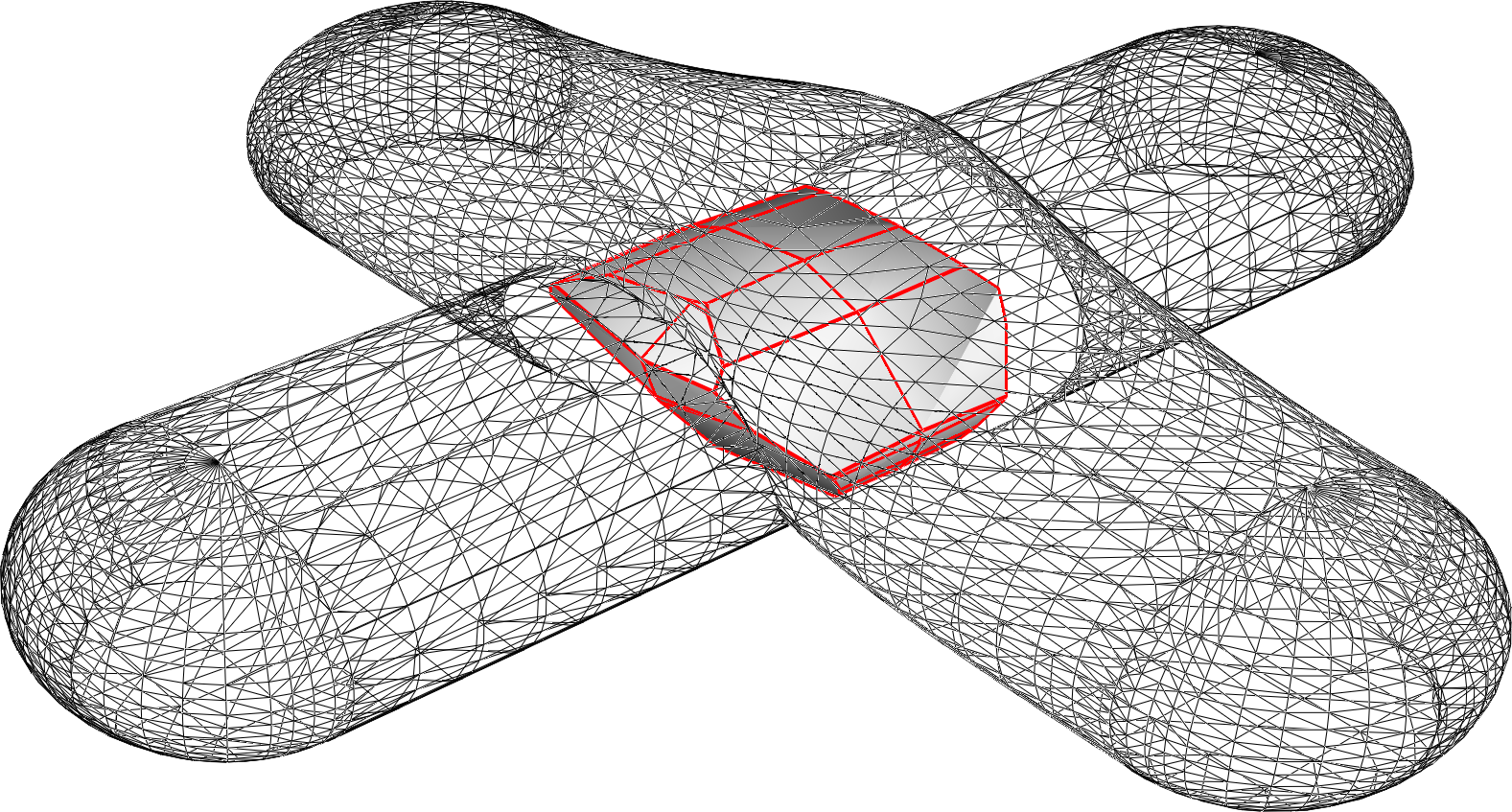} \\
\textit{plus} & \textit{star} & \textit{flex} & \textit{cross} \\
\vspace{0.3cm} \\
\includegraphics[width=.2\linewidth]{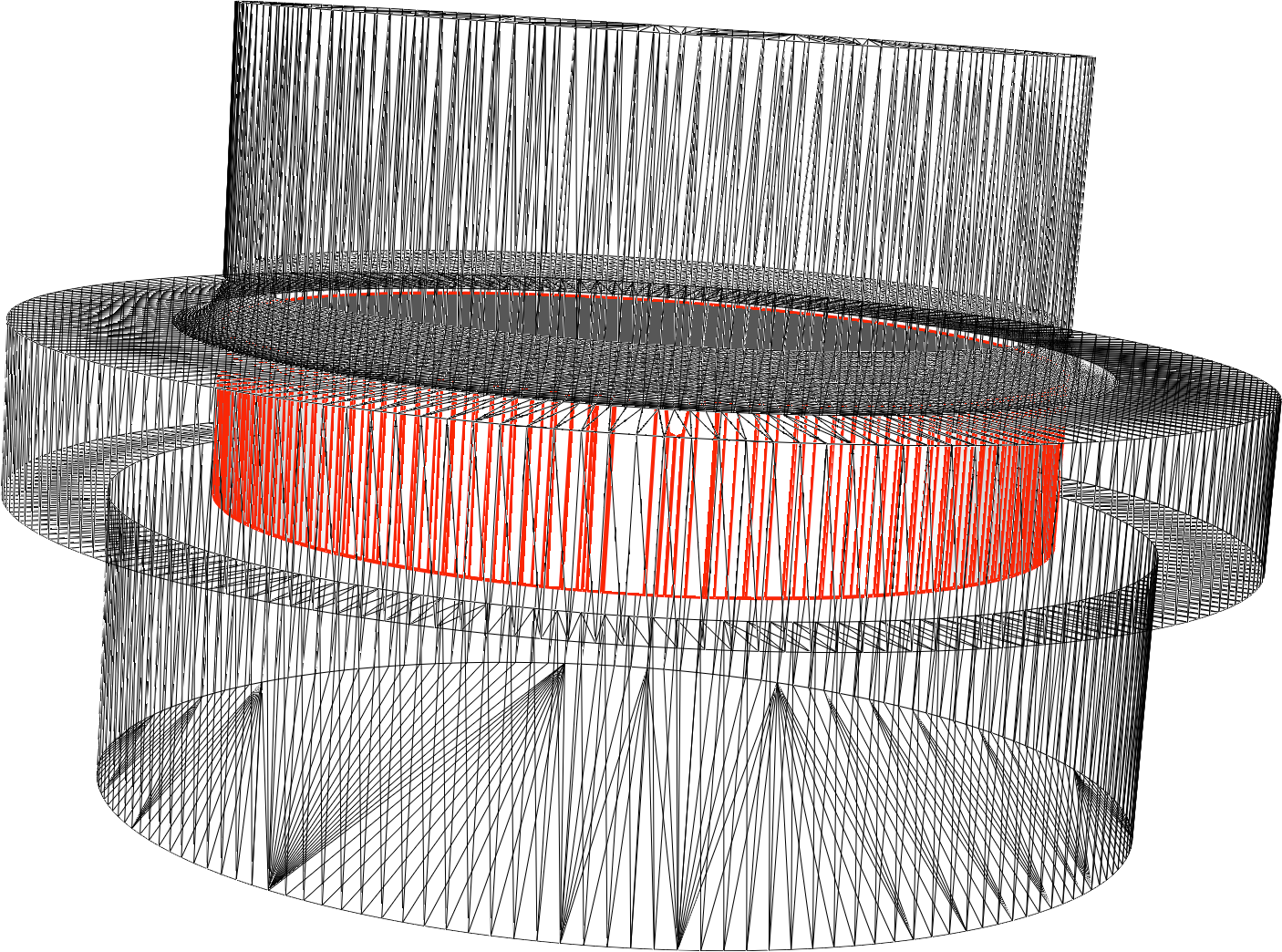} &
\includegraphics[height=.2\linewidth]{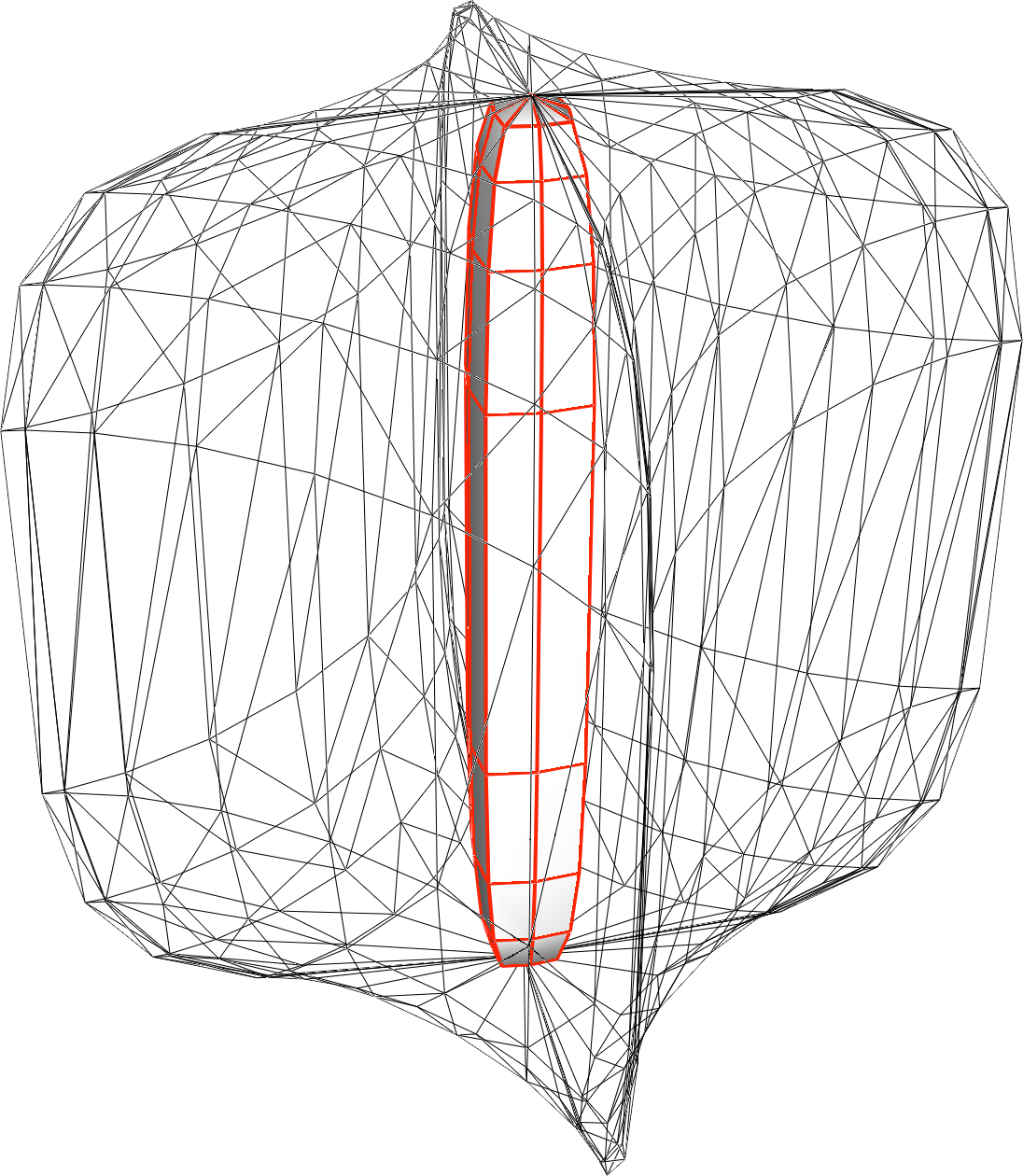} &
\includegraphics[width=.2\linewidth]{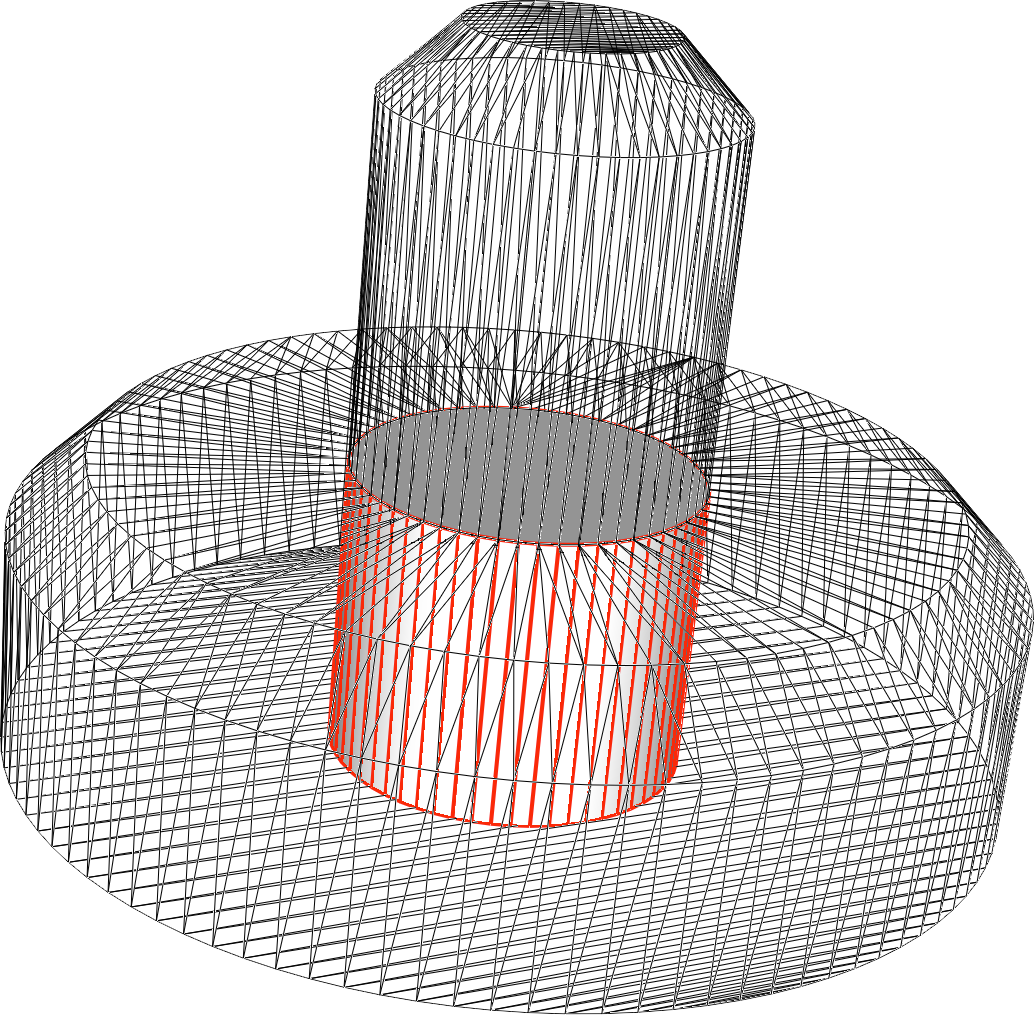} &
\includegraphics[width=.2\linewidth]{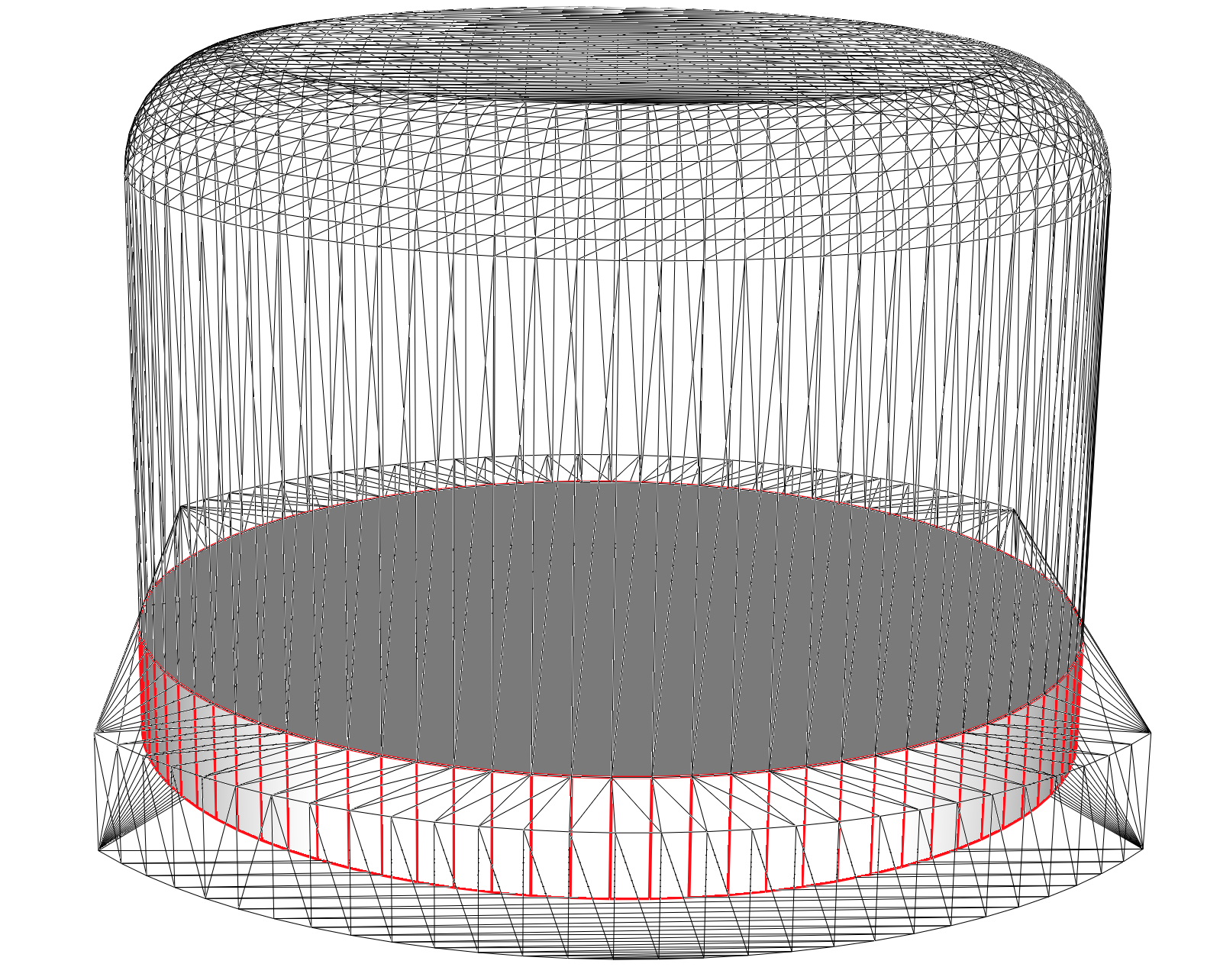} \\
\textit{part} & \textit{super-ellipse} & \textit{bot-eye} & \textit{button} \\
\vspace{0.3cm} \\
\includegraphics[height=.2\linewidth]{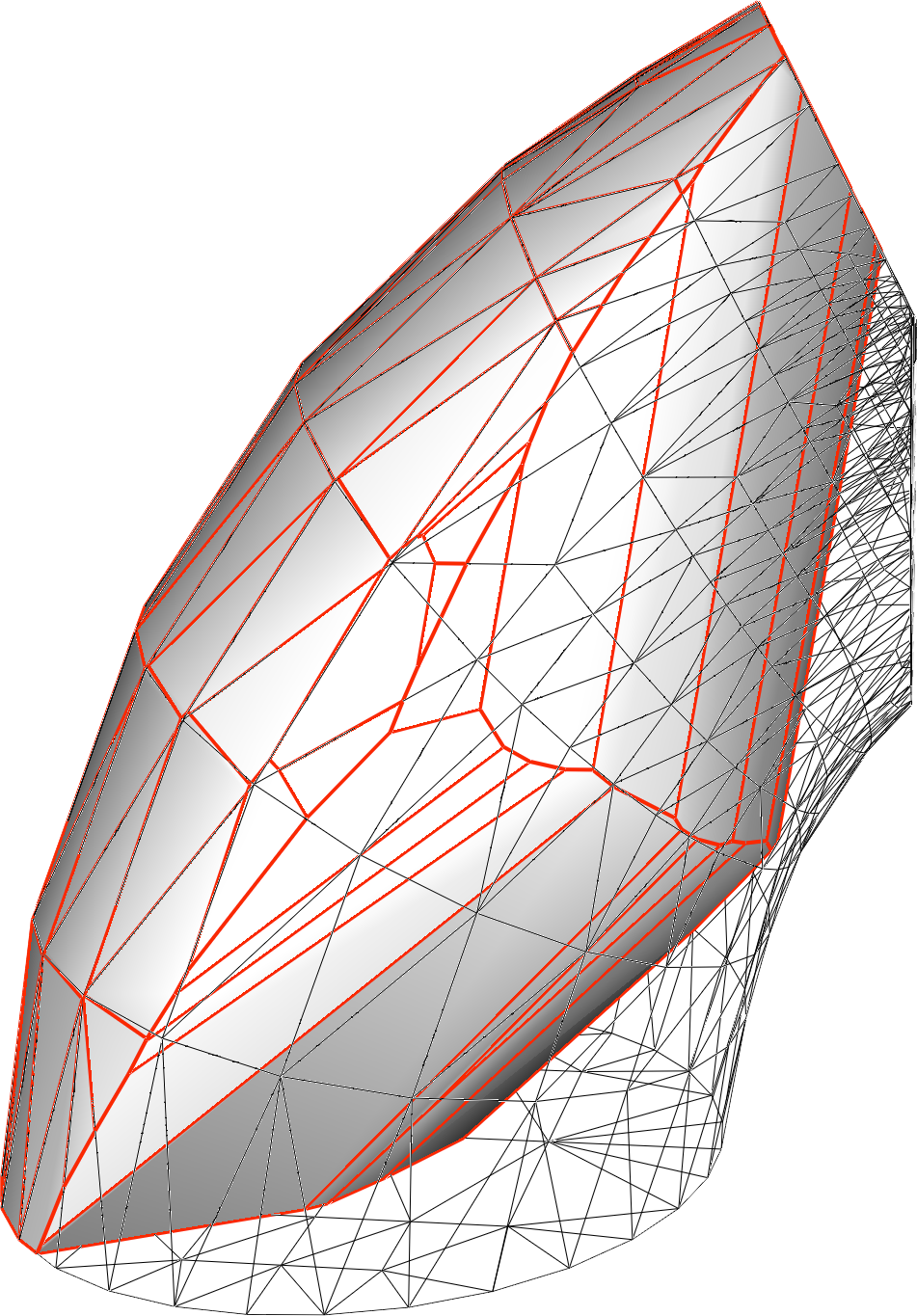} &
\includegraphics[width=.2\linewidth]{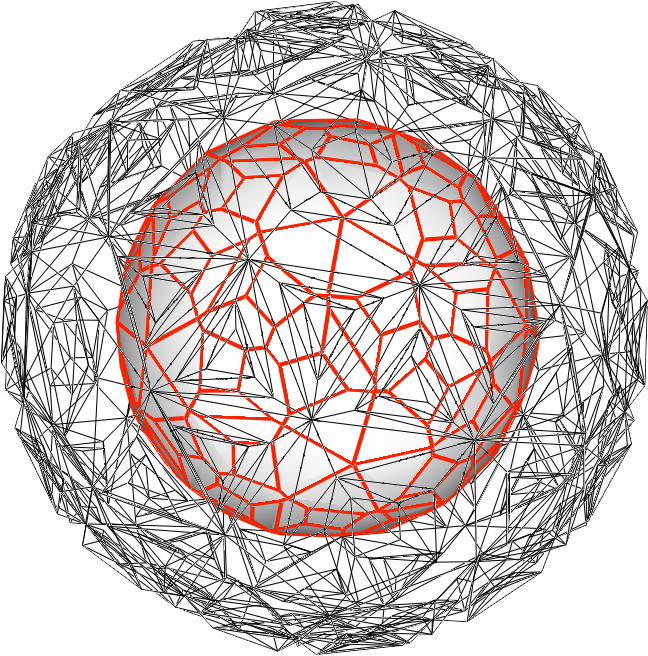} &
\includegraphics[height=.2\linewidth]{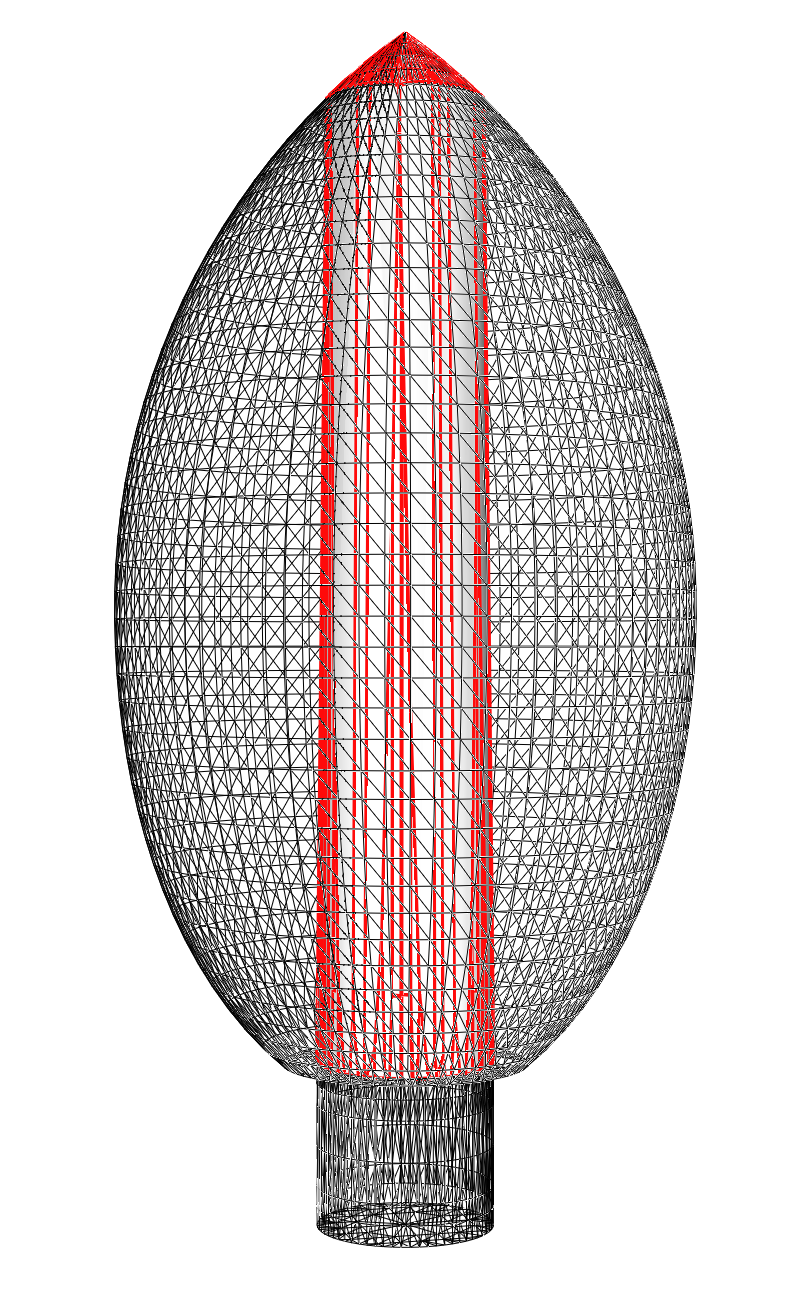} &
\includegraphics[width=.2\linewidth]{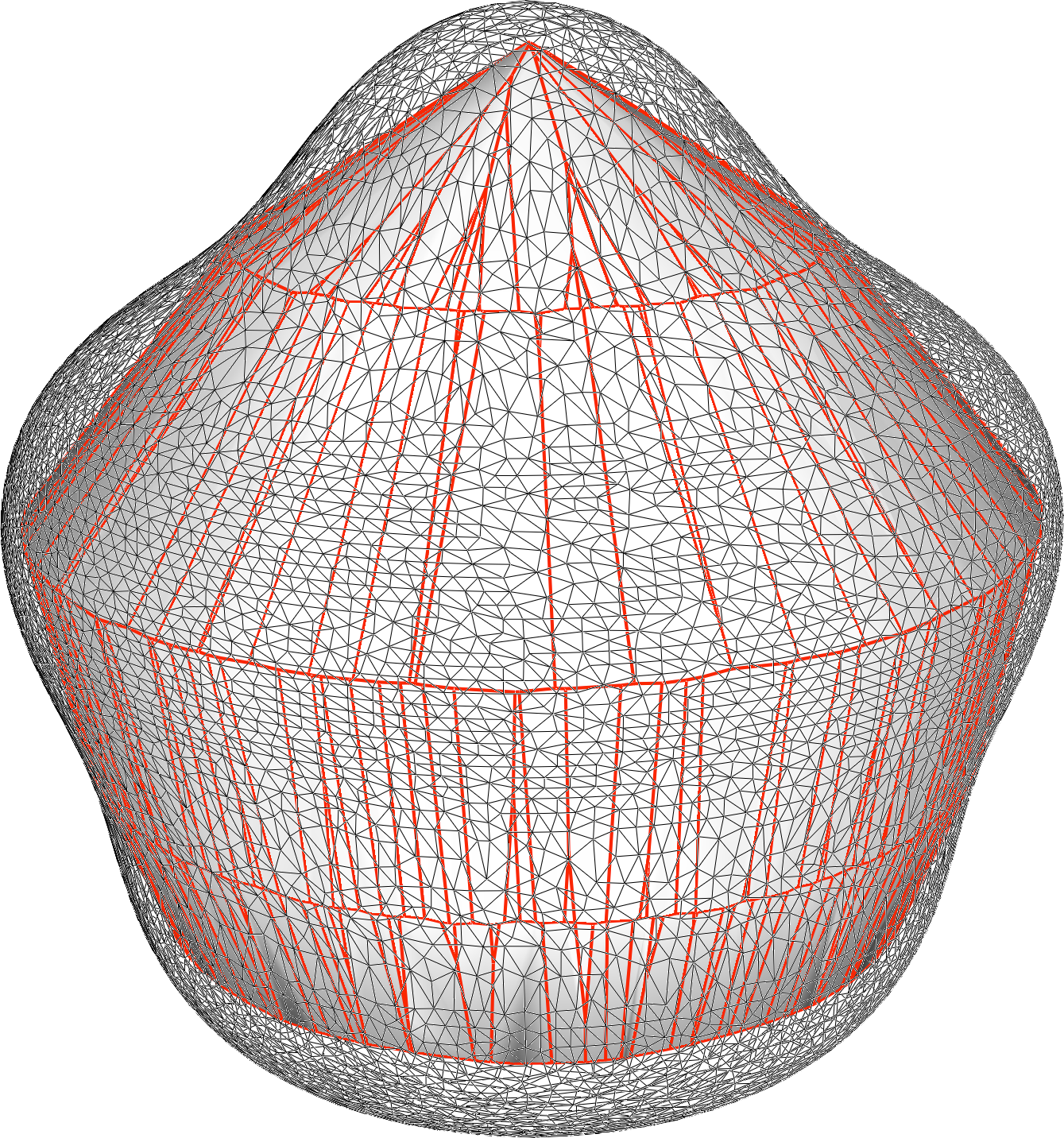} \\
\textit{rt4-arm} & \textit{ball} & \textit{acorn} & \textit{muffin} \\
\end{tabular}
\caption{Examples of our kernel evaluation for complex models.
In the top row models on which the geometric method is more efficient, in the middle models for which the performance are similar and in the bottom row models on which the algebraic method is preferable.}
\label{fig:complex}
\end{figure*}

To further investigate on this point, in Fig.~\ref{fig:complex} we present the kernel computation of 10 selected star-shaped examples from this dataset.
In the top row we have models on which the geometric method is by far more efficient: \textit{plus} (ThingiID: 1120761), \textit{star} (ThingiID: 313883), \textit{flex} (ThingiID: 827640) and \textit{cross} (ThingiID: 313882).
In the middle row, models for which the performance are similar: \textit{part} (ThingiID: 472063), \textit{super-ellipse} (ThingiID: 40172), \textit{bot-eye} (ThingiID: 37276) and \textit{button} (ThingiID: 1329185).
Then in the bottom row we show models on which the algebraic method is preferable: \textit{rt4-arm} (ThingiID: 39353), \textit{ball} (ThingiID: 58238), \textit{acorn} (ThingiID: 815480), \textit{muffin} (ThingiID: 101636). 
The computational times, together with the ones relative to the whole dataset, are reported in Table~\ref{table:time:complex}.

\begin{table}[th]
\caption{Computational times for complex models.
The first number relative to \textit{Thingi} dataset indicates the number of models instead of the number of vertices.}
\label{table:time:complex}
\centering
\begin{tabular}{ccccc}
\hline\noalign{\smallskip}
mesh & $\#$vertices & our-shuffle & CGAL & ratio \\
\noalign{\smallskip}\hline\noalign{\smallskip}
\textit{plus}           & 448 & 0.004 & 0.09 & 22.75\\
\textit{star}           & 9633 & 0.4 & 5.15 & 12.93\\
\textit{flex}           & 834 & 0.02 & 0.27 & 12.76\\
\textit{cross}          & 3914 & 0.19 & 2.1 & 11.12\\
\textit{part}           & 5382 & 2.58 & 6.94 & 2.69\\
\textit{super-ellipse}  & 290 & 0.02 & 0.04 & 2.05\\
\textit{bot-eye}        & 453 & 0.03 & 0.03 & 0.96\\
\textit{button}         & 1227 & 0.1 & 0.08 & 0.75\\
\textit{rt4-arm}        & 655 & 0.13 & 0.09 & 0.67\\
\textit{ball}           & 660 & 0.24 & 0.04 & 0.15\\
\textit{acorn}          & 4114 & 4.35 & 0.55 & 0.13\\
\textit{muffin}         & 8972 & 11.73 & 0.54 & 0.04\\
\textit{\textbf{Thingi dataset}} & 1806 & 172.88 & 518.2 & 2.99\\
\noalign{\smallskip}\hline
\end{tabular}
\end{table}

Once again, we notice that the size of the model impacts on the performance of our method.
Looking at Fig.~\ref{fig:time:thingi} we can see how the models for which our method performs worse than CGAL are all in the right part of the plane, relative to models with a high number of vertices.
At the same time, 
the number of vertices of the element, by itself, is not sufficient to justify the supremacy of one method over the other.
For example, models \textit{star} and \textit{flex} have very different sizes and times, but their ratio is quite similar; the same holds for models \textit{parts} and \textit{super-ellipse} or \textit{ball} and \textit{acorn}.

The shape of the object also plays an important role: over models with numerous adjacent co-planar faces like \textit{plus}, \textit{star} (whose bottom is completely flat) and \textit{cross} our method is preferable even when the size grows.
As already seen in Section~\ref{subsec:refinements}, the presence of coplanar faces significantly improves the performance of our method.
Vice-versa, over elements with significant curvatures like \textit{rt4-arm}, \textit{acorn} or \textit{muffin}, the algebraic method performs similarly or better than ours even on relatively small models like \textit{bot-eye}.
Over these models it is still possible to compute a correct kernel with the geometric approach, but the ratio between CGAL time and ours is in the order of $10^{-1}$ or even $10^{-2}$.

\subsection{Comparison with the previous version}
\label{subsec:comparison}
With respect to its introduction in \cite{SorgenteKernel}, we believe that the algorithm is now more easy to read and to understand, thanks to the introduction of labels for storing the position of vertices, edges and faces with respect to a plane.
Another significant difference is that the evaluation of the position of the vertices is now computed once and for all, at the top level (in Algorithm~\ref{alg:kernel}), while in the previous version every algorithm contained some vertices evaluations: this further reduced the computational complexity.
In addition, we switched from the inexact predicates (in \cite{SorgenteKernel} we used the equivalent of \textit{orient3d-fast}) to the exact \textit{orient3d}, which resulted in an increasing precision in the treatment of nearly-coplanar faces, for a small extra cost which was easily compensated by the other improvements.
This switch required a small modification of the \textit{Plane} class: in view of the usage of \textit{orient3d}, for every plane we also store three points on it.
Last, the introduction of the shuffle mode made the computation of the kernels of non star-shaped objects almost immediate, marking a huge difference with respect to the algebraic method and our previous implementation.
In Table~\ref{table:comparison} we report the differences between the old version of the algorithm and the current version in standard and in shuffle mode.
Over the meshes from Section~\ref{subsec:meshes} there is almost no differences between the three versions.
For refined models, there is an improvement of a factor 2 in the computation of the \textit{vase} refinement (we report the sum of the times for all the refinements).
With complex models we have the greatest differences; in these cases we can also appreciate the advantage brought by the shuffle mode, which was not significant in the other tests.

\begin{table}[htbp]
\caption{Performance comparison between the current implementation and the one presented in \cite{SorgenteKernel}.
Only models with significant differences are reported.}
\label{table:comparison}
\centering
\begin{tabular}{cccc}
\hline\noalign{\smallskip}
mesh & our-shuffle & our & our\cite{SorgenteKernel} \\
\noalign{\smallskip}\hline\noalign{\smallskip}
\textit{spiral} (sum)   & 1.04 & 1.01 & 1.03\\
\textit{vase} (sum)     & 245.61 & 194.91 & 495.63\\
\textit{cross}           & 0.19 & 4.15 & 9.14\\
\textit{part}            & 2.58 & 5.22 & 12.7\\
\textit{bot-eye}         & 0.03 & 0.17 & 0.25\\
\textit{rt4-arm}         & 0.13 & 0.14 & 0.21\\
\textit{button}          & 0.1 & 0.11 & 0.15\\
\textit{ball}            & 0.24 & 0.28 & 0.58\\
\textit{acorn}           & 4.35 & 276.13 & 626.72\\
\textit{muffin}          & 11.73 & 58.62 & 129.38\\
\noalign{\smallskip}\hline
\end{tabular}
\end{table}

\section{Conclusions}
\label{sec:conclusions}
We presented an algorithm for the computation of the kernel of a polyhedron based on the extension to the 3D case of the geometric approach commonly adopted in two dimensions. With respect to \cite{SorgenteKernel} we have optimized the algorithm in several ways: we now perform all the vertices comparisons in a single, preliminary step; we have introduced the use of exact geometric predicates and a random visiting strategy of the faces that considerably improve the performance of the method over non star-shaped, complex objects.
The algorithm showed up to be robust and reliable, as it computed successfully the kernel of every considered polyhedron.
The efficiency of our algorithm is compared to the CGAL implementation of the algebraic approach to the problem.
From a theoretical point of view, the computational complexity evaluation of Section~\ref{subsec:computational_complexity} suggests that our method is in general quadratic, while the algebraic approach has a lower bound at $n\log(n)$.
Nonetheless, we proved all across Section~\ref{sec:tests} that in several circumstances our approach outperforms the algebraic one.

The geometric approach showed up to be significantly faster than the algebraic one when dealing with models satisfying at least one of the following conditions:
\begin{enumerate}
    \item the size of the model is small;
    \item the model contains a significant number of co-planar faces;
    \item the kernel is empty.
\end{enumerate}
Our method performs significantly better than the algebraic approach over polyhedra with a limited number of vertices and faces, as shown in Section~\ref{subsec:meshes}, making it particularly suitable for the analysis of volumetric tessellations with non-convex elements.
Indeed, we point out that our algorithm is specifically designed to be used with simple polyhedra, possibly composing a bigger and more complex 3D model, rather than with a complete model itself.
This behaviour is particularly evident with model \textit{vase} from Section~\ref{subsec:refinements}.
As long as the size of the model remains reasonable our method is faster than CGAL, then over a certain bound the algebraic method becomes more efficient.
Again, models like \textit{super-ellipse} or \textit{flex} from Section~\ref{subsec:complex_models} have few or zero co-planar faces and a significant kernel, but the size of these meshes is small and the geometric approach offers better performance.
According to Fig.~\ref{fig:time:thingi}, a bound on the number of vertices could be possibly set at around $10^3$.
When the size of the polyhedron increases, our method is still particularly efficient if the model has numerous co-planar faces, due for instance to the presence of flat regions on the surface.
This is a very common situation in models representing mechanical parts.
For instance, models \textit{star} and \textit{part} from Section~\ref{subsec:complex_models} present large flat regions despite having a significant size, and again the geometric approach is faster on these models.
Another scenario in which the geometric approach overperforms the algebraic one is with non star-shaped objects.
The differences in this case are so evident that one could even imagine to use our algorithm to understand, in few iterations in shuffle mode, if a model is actually star-shaped or not, even without computing the proper kernel.
On the other side, the algebraic approach is likely to remain preferable over domains which do not satisfy any of the above conditions: star-shaped objects with thousands of vertices and high surface curvature.

In conclusion, with this work we do not aim at completely replacing the algebraic approach for the kernel computation but instead to give an alternative which can be preferred for specific cases, such as the quality analysis of the elements in a 3D tessellation, in the same way as bubble-sort is to be preferred to optimal sorting algorithms when dealing with very small arrays.
As a future development, we plan to integrate in our algorithm the promising \textit{indirect predicates} introduced in \cite{attene2020indirect}.
Numerical problems remain a critical issue in the computation of geometric constructions like the kernel, independently of the approach adopted.
We believe indirect predicates could enormously help in enhancing the robustness of the algorithm.
Moreover, we plan to include this tool in a suite for the generation and analysis of tessellations of three dimensional domains, aimed at PDE simulations.
The kernel of a polyhedron has a great impact on its geometrical quality, and the geometrical quality of the elements of a mesh determines the accuracy and the efficiency of a numerical method over it.
We are therefore already using this algorithm in works like \cite{sorgente2022role} for better understanding the correlations between the shape of the elements and the performance of the numerical simulations, and be able to adaptively generate, refine or fix a tessellation accordingly to them.

\section*{Acknowledgements}
We would like to thank Dr. M. Manzini for the precious discussions and suggestions, and all the people from IMATI institute involved in the CHANGE project.
Special thanks are also given to the anonymous reviewers for their comments and suggestions.

This work has been partially supported by the ERC Advanced Grant CHANGE contract N.694515.

\bibliographystyle{plain}

\begin{thebibliography}{10}

\bibitem{ahn2008geometric}
Hyung~T. Ahn and Mikhail Shashkov.
\newblock Geometric algorithms for {3D} interface reconstruction.
\newblock In {\em Proceedings of the 16th international meshing roundtable},
  pages 405--422. Springer, 2008.

\bibitem{antonietti2022high}
Paola~F Antonietti, Michele Botti, Ilario Mazzieri, and Simone~Nati Poltri.
\newblock A high-order discontinuous galerkin method for the
  poro-elasto-acoustic problem on polygonal and polyhedral grids.
\newblock {\em SIAM Journal on Scientific Computing}, 44(1):B1--B28, 2022.

\bibitem{attene2020indirect}
Marco Attene.
\newblock Indirect predicates for geometric constructions.
\newblock {\em Computer-Aided Design}, 126:102856, 2020.

\bibitem{attene2021benchmark}
Marco Attene, Silvia Biasotti, Silvia Bertoluzza, Daniela Cabiddu, Marco
  Livesu, Giuseppe Patan{\'e}, Micol Pennacchio, Daniele Prada, and Michela
  Spagnuolo.
\newblock Benchmarking the geometrical robustness of a virtual element
  {Poisson} solver.
\newblock {\em Mathematics and Computers in Simulation}, 190:1392--1414, 2021.

\bibitem{baeldung}
Baeldung.
\newblock {Baeldung guides and courses}.
\newblock \url{https://www.baeldung.com/cs/sort-points-clockwise}, 2021.
\newblock Last accessed 2021-09-08.

\bibitem{beirao2013basic}
Lauren\c{c}o Beir{\~a}o~da Veiga, Franco Brezzi, Andrea Cangiani, Gianmarco
  Manzini, Donatella Marini, and Alessandro Russo.
\newblock Basic principles of virtual element methods.
\newblock {\em Mathematical Models and Methods in Applied Sciences},
  23(01):199--214, 2013.

\bibitem{berrone2019parallel}
Stefano Berrone, Stefano Scial{\`o}, and Fabio Vicini.
\newblock Parallel meshing, discretization, and computation of flow in massive
  discrete fracture networks.
\newblock {\em SIAM Journal on Scientific Computing}, 41(4):C317--C338, 2019.

\bibitem{BoostLibrary}
Boost.
\newblock {Boost C++ Libraries}.
\newblock \url{http://www.boost.org/}, 2021.
\newblock Last accessed 2021-09-08.

\bibitem{ciarlet2002finite}
Philippe~G. Ciarlet.
\newblock {\em The finite element method for elliptic problems}.
\newblock SIAM, 2002.

\bibitem{cockburn2012discontinuous}
Bernardo Cockburn, George~E. Karniadakis, and Chi-Wang Shu.
\newblock {\em Discontinuous Galerkin methods: theory, computation and
  applications}, volume~11.
\newblock Springer Science \& Business Media, 2012.

\bibitem{demir2018near}
{\.I}lke Demir, Daniel~G Aliaga, and Bedrich Benes.
\newblock Near-convex decomposition and layering for efficient 3d printing.
\newblock {\em Additive manufacturing}, 21:383--394, 2018.

\bibitem{di2019hybrid}
Daniele~A. Di~Pietro and J{\'e}r{\^o}me Droniou.
\newblock {\em The Hybrid High-Order method for polytopal meshes}, volume~19.
\newblock Springer, 2019.

\bibitem{dupont1980polynomial}
Todd Dupont and Ridgway Scott.
\newblock Polynomial approximation of functions in sobolev spaces.
\newblock {\em Mathematics of Computation}, 34(150):441--463, 1980.

\bibitem{fabri2009cgal}
Andreas Fabri and Sylvain Pion.
\newblock Cgal: The computational geometry algorithms library.
\newblock In {\em Proceedings of the 17th ACM SIGSPATIAL international
  conference on advances in geographic information systems}, pages 538--539,
  2009.

\bibitem{jacobson2017libigl}
Alec Jacobson and Daniele Panozzo.
\newblock Libigl: prototyping geometry processing research in c++.
\newblock In {\em SIGGRAPH Asia 2017 courses}, pages 1--172. 2017.

\bibitem{LeePreparata}
Der-Tsai Lee and Franco~P. Preparata.
\newblock An optimal algorithm for finding the kernel of a polygon.
\newblock {\em J. ACM}, 26(3):415–421, July 1979.

\bibitem{levy2015geogram}
Bruno L{\'e}vy and Alain Filbois.
\newblock Geogram: a library for geometric algorithms.
\newblock {\em /}, 2015.

\bibitem{lipnikov2014mimetic}
Konstantin Lipnikov, Gianmarco Manzini, and Mikhail Shashkov.
\newblock Mimetic finite difference method.
\newblock {\em Journal of Computational Physics}, 257:1163--1227, 2014.

\bibitem{livesu2019cinolib}
Marco Livesu.
\newblock cinolib: a generic programming header only {C}++ library for
  processing polygonal and polyhedral meshes.
\newblock In {\em Transactions on Computational Science XXXIV}, pages 64--76.
  Springer, 2019.

\bibitem{PreparataShamos}
Franco~P. Preparata and Michael~I. Shamos.
\newblock {\em Computational Geometry: An Introduction}.
\newblock Springer-Verlag, Berlin, Heidelberg, 1985.

\bibitem{Brenner-Scott:2008}
Ridgway Scott and Susanne~C. Brenner.
\newblock {\em The mathematical theory of finite element methods}.
\newblock Texts in applied mathematics 15. Springer-Verlag New York, 3 edition,
  2008.

\bibitem{ShamosHoey}
Michael~I. Shamos and Dan Hoey.
\newblock Geometric intersection problems.
\newblock In {\em 17th Annual Symposium on Foundations of Computer Science
  (sfcs 1976)}, pages 208--215, 1976.

\bibitem{shewchuk1997adaptive}
Jonathan~Richard Shewchuk.
\newblock Adaptive precision floating-point arithmetic and fast robust
  geometric predicates.
\newblock {\em Discrete \& Computational Geometry}, 18(3):305--363, 1997.

\bibitem{sorgente2021polyhedral}
T.~Sorgente, S.~Biasotti, G.~Manzini, and M.~Spagnuolo.
\newblock Polyhedral mesh quality indicator for the virtual element method.
\newblock {\em arXiv preprint arXiv:2112.11365}, 2021.

\bibitem{SorgenteKernel}
T.~Sorgente, S.~Biasotti, and M.~Spagnuolo.
\newblock {A Geometric Approach for Computing the Kernel of a Polyhedron}.
\newblock In P.~Frosini, D.~Giorgi, S.~Melzi, and E.~Rodolà, editors, {\em
  Smart Tools and Apps for Graphics - Eurographics Italian Chapter Conference},
  pages 11--19, online, 2021. The Eurographics Association.

\bibitem{sorgente2021vem}
T.~Sorgente, D.~Prada, D.~Cabiddu, S.~Biasotti, G.~Patane, M.~Pennacchio,
  S.~Bertoluzza, G.~Manzini, and M.~Spagnuolo.
\newblock {\em {VEM} and the {M}esh}, volume~31 of {\em SEMA SIMAI Springer
  series}, chapter~1, pages 1--54.
\newblock Springer, 2021.
\newblock ISBN: 978-3-030-95318-8.

\bibitem{sorgente2022role}
Tommaso Sorgente, Silvia Biasotti, Gianmarco Manzini, and Michela Spagnuolo.
\newblock The role of mesh quality and mesh quality indicators in the virtual
  element method.
\newblock {\em Advances in Computational Mathematics}, 48(1):1--34, 2022.

\bibitem{zhou2016thingi10k}
Qingnan Zhou and Alec Jacobson.
\newblock Thingi10k: A dataset of 10,000 {3D}-printing models, 2016.

\end{thebibliography}

\end{document}